\documentclass{jfm}

\usepackage{amsmath}

\usepackage{subfig}
\usepackage{natbib}

\usepackage{hyperref}
\hypersetup{
	breaklinks,
	colorlinks=true,
	linkcolor=blue,
	citecolor=purple,
	pdfauthor={B F Farrell, P J Ioannou, J Jimenez, N C Constantinou, A Lozano-Duran, M-A Nikolaidis},
	pdftitle={A statistical state dynamics based study of large-scale structure in plane Poiseuille flow}
}

\usepackage{microtype}

\usepackage{bm}

\usepackage[usenames,dvipsnames]{xcolor}

\definecolor{ao(english)}{rgb}{0.0, 0.5, 0.0}

\usepackage[normalem]{ulem}

\def\U{\bm{\mathsf{U}}}
\def\Uv{\mathbf{U}}
\def\uv{\mathbf{u}}
\def\kv{\mathbf{k}}

\def\P{{\bf P}}

\def\A{\bm{\mathsf{A}}}

\newcounter{saveeqn}%

\newcommand{\be}{\begin{equation}}
\newcommand{\ee}{\end{equation}}
\newcommand{\bdm}{\begin{equation*}}
\newcommand{\edm}{\end{equation*}}
\newcommand{\bea}{\begin{eqnarray}}
\newcommand{\eea}{\end{eqnarray}}

\newcommand{\partialf}[2]
{
 \ifthenelse{\equal{#1}{}}{\frac{\partial}{\partial #2}}{\frac{\partial #1}{\partial #2}}
}

\newcommand{\sgn}{\mathop{\mathrm{sgn}}}
\newcommand{\real}{\mathop{\mathrm{Re}}}

\renewcommand{\(}{\left(}
\renewcommand{\)}{\right)}
\renewcommand{\[}{\left[}
\renewcommand{\]}{\right]}

\newcommand{\Del}{\Delta}

\newcommand{\df}{\textrm{d}}

\newcommand{\la}{\lambda}

\renewcommand{\i}{\mathrm{i}}

\providecommand\bnabla{\boldsymbol{\nabla}}
\providecommand\bcdot{\boldsymbol{\cdot}}

\usepackage{upgreek}

\def\bit{\vphantom{\dot{W}}}

\renewcommand{\Re}{Re}
\newcommand{\Ret}{Re_\tau}
\newcommand{\ut}{u_\tau}

\renewcommand{\U}{\mathbf{U}}
\renewcommand{\u}{\mathbf{u}}

\renewcommand{\equiv}{\ensuremath{\stackrel{\mathrm{def}}{=}}}
\usepackage{soul}

\begin{document}

\newtheorem{lemma}{Lemma}
\newtheorem{corollary}{Corollary}

\shorttitle{SSD study of large-scale structure in Poiseuille flow} 
\shortauthor{B. F. Farrell and others} 

\title{\vspace{+2ex}A statistical state dynamics-based study  of the structure and mechanism of  large-scale motions in plane Poiseuille flow}

\author
 {
 Brian~F.~Farrell\aff{1},
 Petros~J.~Ioannou\corresp{\email{pjioannou@phys.uoa.gr}}\aff{2},
 Javier~Jim\'enez\aff{3},
 Navid~C.~Constantinou\aff{4},
 Adri\'an~Lozano-Dur\'an\aff{3}
 \and
 Marios-Andreas~Nikolaidis\aff{2}
  }

\affiliation
{
\aff{1}
Department of Earth and Planetary Sciences, Harvard University, Cambridge, MA~02138, USA
\aff{2}
Department of Physics, National and Kapodistrian University of Athens, Panepistimiopolis, Zografos, Athens, 157 84, Greece
\aff{3}
School of Aeronautics, Universidad Polit\'ecnica de Madrid, 28040, Madrid, Spain
\aff{4}
Scripps Institution of Oceanography, University of California San Diego, La Jolla, CA~90293, USA
}

\maketitle

\begin{abstract}
The perspective of statistical state dynamics (SSD) has recently been  applied to the study of
mechanisms underlying turbulence in a variety of
physical systems. An SSD  is a dynamical system that 
evolves a representation of the statistical state of the system. An example  of an SSD 
is the second order  cumulant closure referred to as stochastic structural stability theory (S3T), which has
provided insight into the dynamics of  wall turbulence, and
specifically the emergence and maintenance of the roll/streak structure.
S3T comprises a coupled set of equations for the streamwise mean
and perturbation covariance, in which nonlinear interactions among
the perturbations has been removed, restricting nonlinearity in the dynamics to that of
the mean equation and the interaction between the mean and perturbation covariance.
In this work, this quasi-linear restriction of the dynamics is used  to study the
structure and dynamics of turbulence in plane Poiseuille
flow at moderately high Reynolds
numbers in a closely related dynamical system, referred to as the
restricted non-linear (RNL) system. Simulations using this RNL system reveal that the essential features of wall-turbulence dynamics are retained.
Consistent with previous analyses based on the S3T version of SSD,
the RNL system
spontaneously limits the support of its turbulence to a small set of
streamwise Fourier components giving rise to a naturally minimal
representation of its turbulence dynamics. Although greatly simplified,
this RNL turbulence exhibits natural-looking structures and statistics albeit with quantitative differences from  those  in  direct numerical simulations (DNS)
of the full equations.
Surprisingly, even when further truncation of the perturbation support
to a single streamwise component is imposed,  the RNL system continues to
 self-sustain turbulence with qualitatively realistic structure and dynamic properties.
 RNL turbulence at the Reynolds numbers studied is dominated by the
roll/streak structure in the buffer layer and similar very-large-scale structure
(VLSM) in the outer layer. In this work,  diagnostics of the structure,
spectrum and energetics of RNL and DNS turbulence are used to
demonstrate that the roll/streak dynamics supporting the turbulence in
the buffer and logarithmic layer is essentially similar in RNL and DNS.
\end{abstract}

\section{Introduction}

The fundamental importance of the roll/streak structure in the dynamics of wall-turbulence
was recognized soon after it was observed in the near-wall region in boundary layer flows~\citep{Hama-etal-1957, Kline-etal-1967} and  seen in  early direct numerical simulations (DNS) of channel flows (cf.~\citet{Kim-etal-1987}). Recently, in both experiments and numerical simulations of turbulent flows at high Reynolds numbers, roll/streak structures have been identified in the logarithmic layer with self-similar scale increasing with the distance from the wall~\citep{DelAlamo-etal-2006,Hellstrom-etal-2011,Lozano-Duran-etal-2012,Lozano-Duran-Jimenez-2014,Hellstrom-etal-2016}. Further from the wall similar very large streak structures are seen that scale with the channel half-height or pipe radius, $h$, or with the boundary layer thickness, $\delta$ \citep{Bullock-etal-1978,Jimenez-1998,Kim-Adrian-1999}. These coherent motions are variously referred to as superstructures or very large-scale motions (VLSM) \citep{DelAlamo-etal-2004,Toh-Itano-2005,Hutchins-Marusic-2007,Marusic-etal-2010}.

In this paper the mechanism maintaining these large and very large scale structures in the upper layers of turbulent  plane Poiseuille flow is studied by making use of the methods of  statistical state dynamics (SSD) {with  averaging operator the streamwise average. Conveniently, the fundamental dynamics of wall-turbulence is contained in the simplest nontrivial SSD: a second-order closure of the equations governing the cumulants of the full SSD, referred to as  the stochastic structural stability theory (S3T) system \citep{Farrell-Ioannou-2003-structural,Farrell-Ioannou-2012}. Restriction of the dynamics to the first two cumulants involves either parameterizing the third cumulant by stochastic excitation or, as we will adopt in this work, setting it to zero.  
With the chosen streamwise averaging operator, the first cumulant or mean flow is the streamwise-averaged flow,  i.e. the flow component with streamwise wavenumber $k_x=0$, and the second cumulant is the covariance of the perturbations, which are defined as the deviations from the mean, i.e. the flow components  with wavenumber $k_x\ne 0$. Either stochastic parameterization or setting the third cumulant to zero results in a quasi-linear dynamics in which perturbation-perturbation interactions are not explicitly calculated
~\ref{fig:triads}). We refer to this quasi-linear approximation to the Navier--Stokes equations (NS) used in S3T as the RNL (Restricted Nonlinear) approximation.
The RNL equations are derived from the NS by making the same dynamical restriction as that of  S3T.
A significant consequence of the S3T and RNL restriction to the NS dynamics is the elimination of the classical perturbation--perturbation turbulent cascade. 

This second-order closure has been justified by \cite{Bouchet-etal-2013} in the limit $\lambda \tau\ll1$, 
where $\tau$ is the shear time scale of the large-scale flow and  $1/\lambda$ the relaxation time associated with damping of the large-scale flow. However,  S3T theory is  predictive even when  $\lambda\tau\gg1$.  
For example, it predicts the bifurcation from the statistically homogeneous to the inhomogeneous state that has been confirmed  using DNS both in barotropic turbulence~\citep{Constantinou-etal-2014} and three-dimensional Couette flow turbulence~\citep{Farrell-Ioannou-2016-bifur}.

The S3T closure  had been used to study large-scale
coherent structure dynamics in planetary turbulence~\citep{Farrell-Ioannou-2003-structural,Farrell-Ioannou-2007-structure,Farrell-Ioannou-2008-baroclinic,Farrell-Ioannou-2009-equatorial,Farrell-Ioannou-2009-closure,Marston-etal-2008,Srinivasan-Young-2012,Bakas-Ioannou-2013-prl,Constantinou-etal-2014,Constantinou-etal-2015,Parker-Krommes-2014-generation} and drift-wave turbulence in plasmas~\citep{Farrell-Ioannou-2009-plasmas,Parker-Krommes-2013}. At low  Reynolds numbers the S3T closure of wall-bounded turbulence has been shown to  provide a theoretical framework predicting the emergence and maintenance of the roll/streak structure
in transitional flows  as well as the mechanism sustaining the turbulent flow subsequent to transition.
In particular, S3T  predicts  a new mechanism for the emergence of the roll/streak structure  in pre-transitional flow through a free-stream-turbulence-induced  modal instability
and its equilibration
at finite amplitude.   In addition, S3T  provides a theory for maintenance
of  turbulence  through  a time-dependent parametric instability  process
after transition~\citep{Farrell-Ioannou-2012}.
(Parametric instability  traditionally refers to
the instability of a  linear non-autonomous system,  a  parameter of which  varies periodically in time  (cf. \cite{Drazin-Reid-81}, section 48).
We generalize the concept of parametric instability to include any linear instability that
is inherently caused by the time dependence of the
system. The reason we have adopted the same term to refer to the instability  induced  by time-dependence in both  periodic and non-periodic
systems
is that the same mechanism underlies the instability in these systems,  which is the inherent non-normality of non-commuting time dependent systems
coupled with  the convexity of the exponential function (cf. \cite{Farrell-Ioannou-1996b, Farrell-Ioannou-1999}).)


%
In S3T the perturbation covariance   is  obtained  directly from a  Lyapunov equation which
is equivalent to obtaining the covariance as an ensemble mean of the perturbation covariances obtained from an infinite ensemble of RNL perturbation equations sharing the same mean flow. In the RNL system used in this work, the Reynolds stresses are obtained from the covariance formed using a single member perturbation ensemble.
 Using a single member of the perturbation ensemble to directly calculate the Reynolds stresses bypasses explicit formation of the perturbation covariance and, in this way,
 it has the advantage over S3T that it can be easily implemented at high resolution  in a DNS (cf.~\citet{Constantinou-etal-Madrid-2014,Thomas-etal-2014-sustain}).

S3T has allowed formulation of new theoretical constructs for understanding the dynamics of turbulence, but is
restricted in application to moderate Reynolds numbers by the necessity to solve explicitly for the perturbation covariance.
RNL, by inheriting the structure of S3T,  allows extension of  S3T  methods of analysis to much higher Reynolds numbers.
While the  perturbation variables  that appear
in the RNL  equations are the velocities, for the purposes of  making contact with the theoretical results of S3T analysis  it is important to take the additional conceptual
step of  regarding the perturbation variable of RNL to be the covariance associated with these 
perturbation variables which is an  approximation to the exact  perturbation covariance of S3T.
Taking this perspective underscores and  makes explicit the parallelism between the S3T exact statistical state dynamics and the RNL approximation to it.

In order to maintain turbulence in wall-bounded shear flow, a mechanism is required to transfer
kinetic energy from the external forcing to  the perturbations. Obtaining understanding of this mechanism is a fundamental
challenge in developing a comprehensive theory of turbulence in these flows. If the
mean flow were stationary and inflectional then fast hydrodynamic linear instabilities might plausibly be invoked as responsible for producing this transfer. However, most wall-bounded shear flows
are both rapidly varying and not inflectional, and the mechanism of energy transfer to the perturbations
involves nonlinear processes that exploit linear transient growth arising from the non-normality of the flow dynamics. The problem with sustaining turbulence using transiently growing perturbations is that these perturbations ultimately decay and must be renewed or else the turbulence is not sustained, as is familiar from the study of rapid distortion theory \citep{Kim-Lim-2000}. A mechanism that exploits nonlinearity to maintain the transiently growing perturbations we refer to as a self-sustaining process (SSP)
\citep{Hamilton-etal-1995,Jimenez-Pinelli-1999,Jimenez-2013}.
The various SSP mechanisms that have been proposed
have in common this sustaining of transient growth associated with the non-normality of the mean flow by
exploiting nonlinearity to renew the transiently growing set of perturbations. Consistent with the roll/streak structure being the structure of optimal linear growth, proposed SSP mechanisms may be regarded as alternative nonlinear processes for renewing this structure \citep{Hamilton-etal-1995,Hall-Sherwin-2010,Jimenez-2013}. In the case of the S3T, and also in the dynamically similar
RNL studied in this work, the mechanism effecting this transfer is known: it is systematic correlation by the streak of perturbation Reynolds stresses that force the roll, which in turn maintains the streak through the lift-up process. These Reynolds stresses are produced by the Lyapunov vectors arising from parametric instability of the time dependent streak. This mechanism, which has analytical expression in S3T and in the noise-free RNL, comprises a cooperative interaction between the coherent streamwise mean flow and the incoherent turbulent perturbations. It was analyzed first using S3T,
and implications of this analysis, including the predicted parametric streak instability and its associated Lyapunov spectrum of modes, have been studied in subsequent work \citep{Farrell-Ioannou-2012,Constantinou-etal-Madrid-2014,Thomas-etal-2014-sustain,Thomas-etal-2015-minimal,Farrell-Ioannou-2016-sync,Nikolaidis-etal-Madrid-2016}. This analysis of the SSP using S3T,  including the parametric instability of the streak and the structure of the perturbation field arising as the leading Lyapunov vector of this parametric growth, will be frequently invoked in this work.
Detailed explanation of these concepts can
be found in the above references.

In this paper, we compare DNS and RNL simulations made without any explicit stochastic parameterization
at relatively high Reynolds numbers in pressure-driven channel flow.
Included in this comparison are flow statistics, structures, and dynamical diagnostics. This  comparison allows the effects of the dynamical restriction in  RNL
to be studied. We find that  RNL with the stochastic parametrization set to zero spontaneously limits the support of its turbulence to a small set of streamwise Fourier
components, giving  rise naturally to
a minimal representation of its turbulence dynamics. Furthermore, the highly simplified RNL dynamics
supports a self-sustaining roll/streak SSP in the buffer layer consistent with that predicted by S3T at
lower Reynolds number and similar to that of DNS. Finally,  we find
that roll/streak structures in the log-layer
are also supported by an essentially similar SSP.

\section{RNL Dynamics}
\label{sec:framework}

Consider a plane Poiseuille flow in which a constant mass flux is maintained by application of a time-dependent pressure, $-G(t)x$, where $x$ is the streamwise coordinate. The  wall-normal direction is $y$ and the spanwise direction is $z$. The  lengths of the channel in the streamwise, wall-normal and spanwise  direction are respectively $L_x$, $2h$ and $L_z$. The channel walls are at $y/h=0$~and~$2$. Averages are denoted by square brackets with a subscript denoting the variable which is averaged, e.g.~spanwise averages by $[\,\bcdot\,]_z=L_z^{-1} \int_0^{L_z} \bcdot\ \df z$, time averages by~$[\,\bcdot\,]_t=T^{-1} \int_0^{T} \bcdot\ \df t$, with  $T$ sufficiently long.
Multiple subscripts denote an average over the subscripted variables  in  the order they appear, e.g. $[\,\bcdot\,]_{x,y} \equiv \[\,[\,\bcdot\,]_{x}\,\]_{y}$. The velocity, $\uv$, is decomposed into its streamwise mean value, denoted by $\Uv(y,z,t)\equiv\[\u(x,y,z,t)\]_x$, and the deviation from the mean (the perturbation), $\u'(x,y,z,t)$, so that  $\uv = \Uv + \uv'$. The pressure
 is similarly written as  $p= -G(t)  x + P(y,z,t)+ p'(x,y,z,t) $.
The NS decomposed into an equation for the  mean and an equation for the perturbation are:
\begin{subequations}
\label{eq:NS}
\begin{gather}
\partial_t\mathbf{U}+ \mathbf{U} \bcdot \bnabla \mathbf{U}  - G(t) \hat{\mathbf{x}} + \bnabla P - \nu \Delta \mathbf{U} = - \[\mathbf{u}' \bcdot \bnabla \mathbf{u}'\]_x\ ,
\label{eq:NSm}\\
 \partial_t\mathbf{u}'+   \mathbf{U} \bcdot \bnabla \mathbf{u}' +
\mathbf{u}' \bcdot \bnabla \mathbf{U}  + \bnabla p' -  \nu \Delta  \mathbf{u}'
= - \(  \mathbf{u}' \bcdot \bnabla \mathbf{u}' - \[\mathbf{u}' \bcdot \bnabla \mathbf{u}'\]_x \,\)\ ,
 \label{eq:NSp}\\
 \bnabla \bcdot \mathbf{U} = 0\ ,\ \ \ \bnabla \bcdot \mathbf{u}' = 0\ ,\label{eq:NSdiv0}
\end{gather}\label{eq:NSE0}\end{subequations}
where $\nu$ is the coefficient of kinematic viscosity.
No-slip boundary conditions are applied in the wall-normal direction and periodic boundary conditions in the streamwise
and spanwise directions.
All nonlinear interactions among the  mean flow components (flow components with streamwise wavenumber $k_x=0$) and perturbations (flow components with streamwise wavenumber $k_x \ne 0$) in~\eqref{eq:NS} are summarized in figure~\ref{fig:triads}.

\begin{figure}
\begin{center}
\centerline{\includegraphics[width=5.3in]{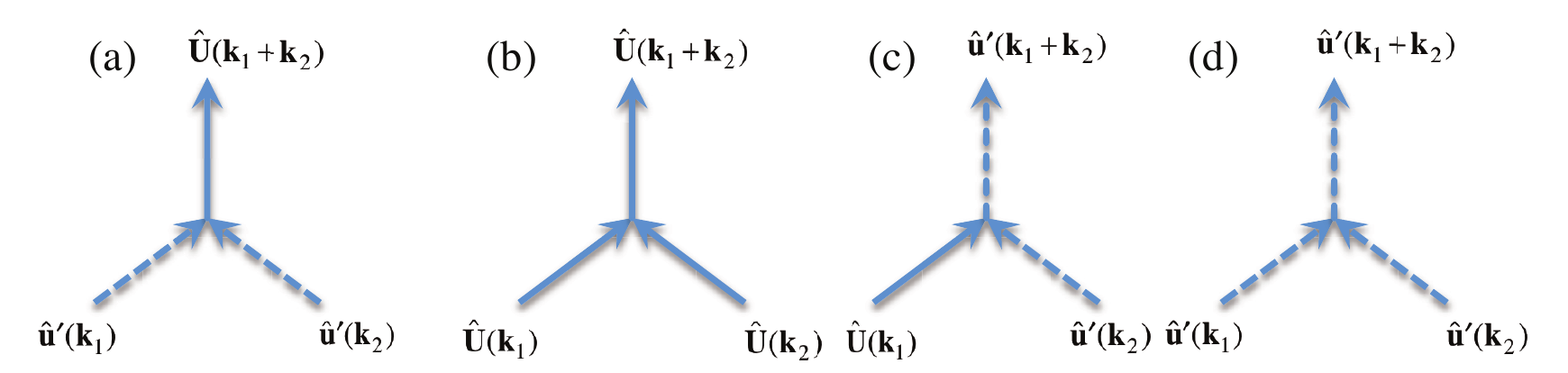}}
\end{center}
\caption{Nonlinear interactions
that are included or excluded in the  S3T and RNL approximations of the NS. Mean flow Fourier components $\hat{\Uv}(\kv)$ with wavenumber
$\kv=(0,k_y,k_z)$ are indicated with solid arrows, the perturbation
Fourier components $\hat{\uv}'(\kv)$ with $k_x \ne 0$ with dashed
arrows. The possible nonlinear interactions are: (a)~a perturbation with
streamwise wavenumber $k_{x_1}$ interacts with another perturbation with
$k_{x_2}=-k_{x_1}$ to produce a mean flow component with $k_x=0$, (b) two
mean flow components interact to make another mean flow component, (c) a
mean flow component interacts with a perturbation to make a perturbation
component and (d) two perturbation components with streamwise
wavenumbers $k_{x_1} \ne - k_{x_2}$ interact to make another perturbation
component. All interactions are included in the NS
equations. Interactions (a) and (b) are included in
the  S3T and RNL mean equations,
while in the S3T and RNL perturbation equations   interactions (c)  are included and interactions (d) are either neglected or stochastically parameterized.\label{fig:triads}}
\end{figure}

The $x,y,z$ components of $\U$ are $(U,V,W)$ and the corresponding components of $\u'$ are $(u',v',w')$.
Streamwise mean perturbation Reynolds stress components are denoted as e.g. $\[u'u'\]_x$, $\[u'v'\]_x$. The streak component of the streamwise  mean flow is denoted by $U_s$ and defined as
\begin{equation}
U_s\equiv U-[U]_z\ .
\end{equation}
The $V$ and $W$  are the streamwise mean velocities of the roll vortices.
We also define the streak  energy density, $E_s=h^{-1}\int_0^h \df y\,\frac1{2}\[U_s^{2}\]_z$, and  the roll energy density,
$E_r=h ^{-1}\int_0^h\df y\,\frac1{2}\[V^2+W^2\]_z$.

The RNL approximation is obtained by neglecting or parameterizing stochastically the perturbation--perturbation interaction terms in~\eqref{eq:NSp} (cf.~figure~\ref{fig:triads}). In this
paper we set the stochastic parameterization  to zero and the RNL system becomes:\begin{subequations}
\label{eq:QL}
\begin{gather}
\partial_t\U+ \U \bcdot \bnabla \U  - G(t) \hat{\mathbf{x}} + \bnabla P - \nu\Delta \U = - \[\u' \bcdot \bnabla \u'\]_x\ ,\label{eq:QLm}\\
\partial_t\u'+   \U \bcdot \bnabla \u' +
\u' \bcdot \bnabla \U  + \bnabla p' -  \nu \Delta  \u'
= 0\ , \label{eq:QLp}\\
\bnabla \bcdot \mathbf{U} = 0\ ,\ \ \ \bnabla \bcdot \mathbf{u}' = 0\ .\label{eq:NSdiv0}
\end{gather}\label{eq:QLE0}\end{subequations}
Equation~\eqref{eq:QLm} describes the dynamics of the streamwise mean flow, $\U$,  which is driven by the divergence of the streamwise mean perturbation Reynolds stresses.
These Reynolds stresses are obtained from~\eqref{eq:QLp} in which
the streamwise-varying perturbations,  $\u'$,  evolve under the  influence of the time dependent streamwise mean flow $\U(y,z,t)$ with no explicitly retained interaction among these streamwise-varying perturbations (the retained interactions are shown in the diagram of~figure~\ref{fig:triads}).
Remarkably, RNL self-sustains turbulence solely due to the perturbation
Reynolds stress forcing of the streamwise mean flow~\eqref{eq:QLm}, in the absence of which a self-sustained turbulent state cannot be established~\citep{Gayme-2010-thesis,Gayme-etal-2010}.

Because the RNL equations do not include interactions among the perturbations,
and because $\Uv$ is streamwise constant, each component $\hat{\uv}'_{k_x}e^{\i k_x x}$ of  perturbation velocity  $\uv'$ in the Fourier expansion:
\begin{equation}
\uv' (x,y,z,t)= \sum_{k_x\in K_x} \real{\[ \bit\hat{\uv}'_{k_x} (y,z,t)\,e^{\i k_x x}\]}\ ,
\end{equation}
where $K_x\equiv 2 \upi/L_x [1,2,3,\dots, N_x/2]$ is all the positive $k_x$ wavenumbers included in the simulation, evolves independently in  equations \eqref{eq:QLp} and therefore \eqref{eq:QLp} can be split into independent equations for each $k_x$. By taking the Fourier transform of \eqref{eq:QLp}
in $x$ and eliminating the perturbation pressure \eqref{eq:QLp}
can be symbolically written as:
 \begin{equation}
 \partial_t \hat{\uv}'_{k_x} = \A_{k_x}(\U)\, \hat{\uv}'_{k_x}\ ,
 \label{eq:Alyap0}
 \end{equation}
with
\begin{equation}
\A_{k_x}(\U) \,\hat{\uv}'_{k_x}={\mathbf P_{L }} \left ( -\Uv \bcdot \bnabla_{k_x} \hat{\uv}'_{k_x} -  \hat{\uv}'_{k_x} \bcdot \bnabla_{k_x} \Uv  + \nu \Delta_{k_x}  \hat{\uv}'_{k_x} \right )\ ,\label{eq:APL}
\end{equation}
and ${\mathbf P_L}$ is the Leray projection enforcing non-divergence of the $k_x$ Fourier components of the perturbation velocity field with $\bnabla_{k_x}\equiv(\i k_x,\partial_y,\partial_z)$ and $\Delta_{k_x}\equiv\partial^2_y+\partial^2_z-k_x^2$ \citep{Foias-etal-2001}. The RNL system can then be written in the form: \begin{subequations}
\begin{gather}
\partial_t\U+ \U \bcdot \bnabla \U  - G(t) \hat{\mathbf{x}} + \bnabla P - \nu\Delta \U = -
\frac{1}{2} \sum_{k_x\in K_x} \real \left [  \bit  \partial_y ( \hat{v}'_{k_x} \hat{\uv}'^{*}_{k_x}) + \partial_z ( \hat{w}'_{k_x} {\hat{\uv}}'^*_{k_x}  )\right ] \ ,
\label{eq:QLmf}\\
\partial_t \hat{\uv}'_{k_x} = \A_{k_x}(\U)\, \hat{\uv}'_{k_x}\ , \ \ {k_x \in K_x}\ , \label{eq:QLpf}\\
\bnabla \bcdot \mathbf{U} = 0\ ,\ \ \ \bnabla_{k_x} \bcdot \hat{\mathbf{u}}'_{k_x} = 0  \ , \label{eq:QLdiv0f}
\end{gather}\label{eq:QLE0f}\end{subequations}
with $*$ in \eqref{eq:QLmf} denoting complex conjugation.

\section{DNS and RNL Simulations}

The data were obtained from a DNS of~\eqref{eq:NSE0} and from an RNL simulation of~\eqref{eq:QLE0}, that is directly associated with the DNS
through eliminating the perturbation--perturbation interaction in the perturbation equation. The dynamics were expressed in the form of
evolution equations for the wall-normal vorticity and the Laplacian of
the wall-normal velocity, with spatial discretization and Fourier
dealiasing in the two wall-parallel directions and Chebychev polynomials
in the wall-normal direction~\citep{Kim-etal-1987}. Time stepping was
implemented using the third-order semi-implicit Runge-Kutta method.

\begin{table}
  \begin{center}
\def~{\hphantom{0}}
\begin{tabular}{@{}*{5}{c}}
\break
 Abbreviation  &$[L_x,L_z]/h$&$N_x\times N_z\times N_y$& $Re_\tau$ &$[L_x^+,L_z^+]$\\
\\
NS940   ~&~ $[\upi\;,\;\upi/2]$ ~&~ $256\times 255\times 385$ ~&~ 939.9   ~&~ $[2953,1476]$\\
RNL940  ~&~  $[\upi\;,\;\upi/2]$ ~&~ $256\times  255\times 385$  ~&~ 882.4 ~&~ $[2772,1386]$\\
RNL940$k_x$12  ~&~  $[\upi\;,\;\upi/2]$ ~&~ $ ~~3 \times  255\times 385$  ~&~ 970.2 ~&~ $[3048,1524]$
\end{tabular}
\caption{Simulation parameters. $[L_x,L_z]/h$ is
the domain size in the streamwise and spanwise direction. $N_x$, $N_z$
are the number of Fourier components after dealiasing and $N_y$ is the
number of Chebyshev components. $\Ret$  is the Reynolds number of the
simulation based on the friction velocity and $[L_x^+,L_z^+]$ is the
channel size in wall units. The Reynolds number based on the bulk velocity is $Re= U_b h/\nu=18511$ in all cases.\label{table:geometry}}
\end{center}
\end{table}

The geometry and resolution of the DNS and RNL simulations is given in
Table~\ref{table:geometry}.

Quantities reported in outer units have lengths  scaled by the channel
half-width, $h$, and time by $h/\ut$ and the corresponding Reynolds
number is $\Ret= \ut h / \nu$ where $\ut= \sqrt{ \nu \left.\df U/\df y\right|_{\rm w}}$
(with $\left.\df U/\df y\right|_{\rm w}$ being the shear at the wall) is the friction
velocity. Quantities reported in inner units have lengths scaled by $h_{\tau} = \Ret^{-1} h$ and
time by $\Ret^{-1} h/\ut$.  Velocities scaled by  the friction velocity $\ut$ will
be denoted with the superscript $+$, which indicates inner unit scaling.

We report results from three simulations: a DNS simulation, denoted NS940, with $\Re_\tau \approx 940$, the corresponding RNL simulation, denoted RNL940,  and
a constrained RNL simulation, denoted RNL940$k_x$12. 
 Both  RNL simulations were initialized with an NS940 state and run until a steady state was established. In RNL940$k_x$12, only the single streamwise Fourier component  with wavenumber $k_x h = 12$ was retained in \eqref{eq:QLpf} by limiting the spectral components of  the perturbation equation to only this streamwise wavenumber; this simulation self-sustained a turbulent state with $\Ret=970.2$. In RNL940, the number of streamwise Fourier components was not constrained; this simulation self-sustained a turbulent state at $\Ret=882.2$.


\begin{figure}
\begin{center}
\centerline{\includegraphics[width=4.8in,trim = 0mm 6mm 0mm 1mm, clip]{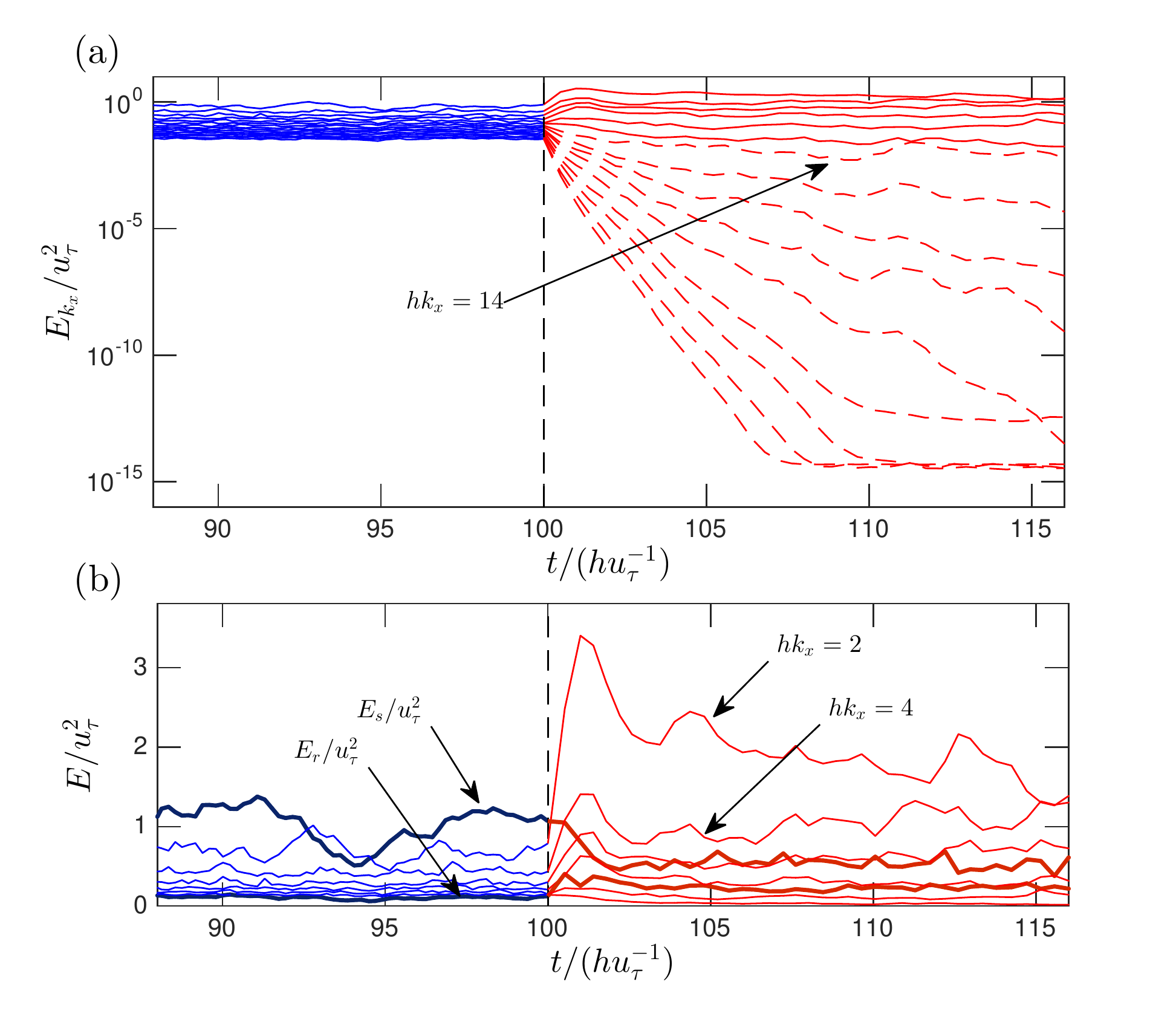}}
\end{center}
	\caption{An NS940 simulation up to $t/(h\ut^{-1})=100$ (indicated with the vertical line) is continued subsequently under RNL dynamics.
	(a)~The energy density, $E_{k_x}$, of the first 15 streamwise varying Fourier components ($h k_x=2,4,\dots,30$). The energy density of the Fourier components decreases monotonically with increasing wavenumber. Decaying Fourier components are indicated with dashed lines. After the transition to RNL dynamics all components with $h k_x \ge 14$  decay ($h k_x=14$  decays, although it is not apparent until later times than shown in this figure). Asymptotically the dynamics of the RNL940 turbulence is maintained by interaction between the set of surviving $h k_x=2,4,\dots,12$ Fourier components and the mean flow ($k_x=0$).
	(b)~Detailed view showing the  energy density of the mean and surviving perturbation components during the transition from NS to RNL dynamics, in which the total energy increased by $10\%$. For the $k_x=0$ shown are: the streak  energy density, $E_s$, and roll energy density, $E_r$. The energy density of the $h k_x=2,4,6,8$ components increases rapidly during the adjustment after transition to RNL dynamics. Note that the total energy density in the perturbation $k_x\ne 0$ components decreases from $0.91u_\tau^2$ in the NS940 ($0.56u_\tau^2$ being in the components that survive in the RNL) to $0.78u_\tau^2$ in RNL940. Also the roll/streak energy density decreases from  $1.1u_\tau^2$ in NS940 to $0.8u_\tau^2$ in RNL940, while the energy density of the $k_x=k_z=0$ component increases from $397u_\tau^2$ to $448u_\tau^2$.\label{fig:Enkx_DNSQL950}}
\end{figure}

We show in figure~\ref{fig:Enkx_DNSQL950} the transition from NS940 to RNL940 turbulence.
The NS940  is switched at time $t u_\tau/ h =100$ to an RNL simulation
by suppressing  the perturbation-perturbation interactions, represented by the right-hand side in equation \eqref{eq:NSp}.
The transition from DNS to RNL is evident in the time series of the energy density
of the streamwise Fourier components of
the perturbation field, given by:
\begin{equation}
E_{k_x}=\frac1{4h} \int_0^{h}\df y\, \[\left| \hat{\uv}'_{k_x} \right|^2\]_{z}~.
\end{equation}
The time evolution of the energy density of the  first 15 streamwise  Fourier components, with wavenumbers $h k_x=2,4,\dots,30$, in NS940 and  in RNL940 is shown in figure~\ref{fig:Enkx_DNSQL950}a. In  NS940,  all $k_x$ components maintain non-zero energy density.
After the transition, the RNL940 turbulence is
maintained by interaction between the set of six surviving wavenumbers, $h k_x=2,4,\dots,12$  and the $k_x=0$ Fourier component of the flow (cf. figure~\ref{fig:Enkx_DNSQL950}a).
The result of  restriction of NS dynamics  to RNL is a spontaneous
reduction in the support of the turbulence in streamwise Fourier components, with all Fourier
components having wavelength smaller than $\upi h/6$ ($h k_x >12$)  decaying exponentially,
producing a reduced complexity dynamics in which turbulence self-sustains on
this greatly restricted support in streamwise Fourier components. We view this transition of NS940 turbulence to RNL940 turbulence as revealing the set of structures that are naturally involved in
maintaining the turbulent state. Given this spontaneous complexity reduction, the question arises:
how few streamwise-varying perturbation components are required in order to self-sustain RNL turbulence
at this Reynolds number? We show in RNL940$k_x$12, that even if we retain only the single
perturbation component with wavelength $\upi h/6$ ($h k_x  =12$), a realistic self-sustained turbulent state persists. (For a discussion of the streamwise wavenumber support of RNL turbulence cf.~\cite{Thomas-etal-2015-minimal}.)

\section{RNL as a minimal turbulence model}
\label{sec:numerics}

We have seen that as a result of its dynamical restriction,
RNL turbulence with the stochastic parameterization set to zero is supported by a small subset of streamwise
Fourier components.  In order to understand this property of RNL dynamics consider that the time dependent streamwise mean state of a turbulent RNL simulation has been stored, so that the mean flow field
$\U(y,z,t)$  is known at each instant.
Then each $k_x$  component of the perturbation flow field  that is retained in the RNL  evolves according to~\eqref{eq:QLpf}:
 \begin{equation}
 \partial_t \hat{\uv}'_{k_x} = \A_{k_x}(\U)\, \hat{\uv}'_{k_x}\ ,\label{eq:Alyap}
 \end{equation}
with $\A_{k_x}(\U)$ given by \eqref{eq:APL}.

With the time dependent mean flow velocity $\U$ obtained  from a
simulation of a turbulent state imposed, equations \eqref{eq:Alyap} are
time dependent linear equations for $\hat{\uv}'_{k_x}$
with the property that each $k_x$ streamwise component of the
perturbation state of the RNL, $\hat{\uv}'_{k_x}$, can be recovered with exponential
accuracy (within an amplitude factor and a phase) by integrating forward \eqref{eq:Alyap} regardless
of the initial state.
This follows from the fundamental property of time dependent systems that all  initial states, $\hat{\uv}'_{k_x}(y,z,t=0)$, converge eventually with exponential accuracy to the same structure  (for a proof cf.~\cite{Farrell-Ioannou-1996b}).
This is completely analogous to the familiar result that regardless of the initial perturbation (with measure zero exception) an autonomous linear system converges to a rank-one structure as time increases: the eigenvector of maximum growth.
In fact, each of the $\hat{\uv}'_{k_x}$ assumes the unique structure of the top Lyapunov vector associated with the maximum Lyapunov exponent of~\eqref{eq:Alyap} at wavenumber $k_x$, which can be obtained by calculating the limit:
\begin{equation}
	{\lambda}_{k_x} = \limsup_{t \rightarrow \infty} \frac{\log \|\hat{\uv}'_{k_x}(y,z,t)\|}{t}~,
\end{equation}
where  $\|\,\bcdot\,\|$ is any norm of the velocity field. 
Moreover,  for each $k_x$, this top Lyapunov exponent has the further property
of being either exactly zero or negative, with those structures having ${\lambda}_{k_x} = 0$
supporting the perturbation variance. The vanishing of the maximum Lyapunov exponents of  linear equations \eqref{eq:Alyap}
reflects the property that  the perturbation $k_x$ components  that are self-sustained in the turbulent state  neither  decay to zero
nor grow without bound.


This property of RNL turbulence being sustained by the top Lyapunov perturbation structures implies that the perturbation structure contains only the streamwise-varying perturbation Fourier components, $k_x$, that are contained in the support of these top Lyapunov structures with ${\lambda}_{k_x} =0$. It is remarkable that only six Fourier components, $k_x$, are
contained in the support of RNL940,  and even more remarkable that the RNL SSP persists even when this naturally reduced set is further truncated to  a single streamwise Fourier component, as demonstrated in RNL940$k_x$12.
This result was first obtained in the case of self-sustained Couette turbulence at low Reynolds numbers  (cf.~\citet{Farrell-Ioannou-2012,Thomas-etal-2014-sustain}).

This vanishing of the   Lyapunov exponent associated with each streamwise wavenumber is enforced in RNL by the nonlinear feedback process acting between the streaks and the perturbations, by which the parametric instability of the perturbations is suppressed at sufficiently high streak amplitude so that the instability in the asymptotic limit  maintains zero Lyapunov exponent \citep{Farrell-Ioannou-2012,Farrell-Ioannou-2016-sync}.

\begin{figure}
	\centering
	\centerline{\includegraphics[width=.53\textwidth,trim = 8mm 0mm 15mm 5mm, clip]{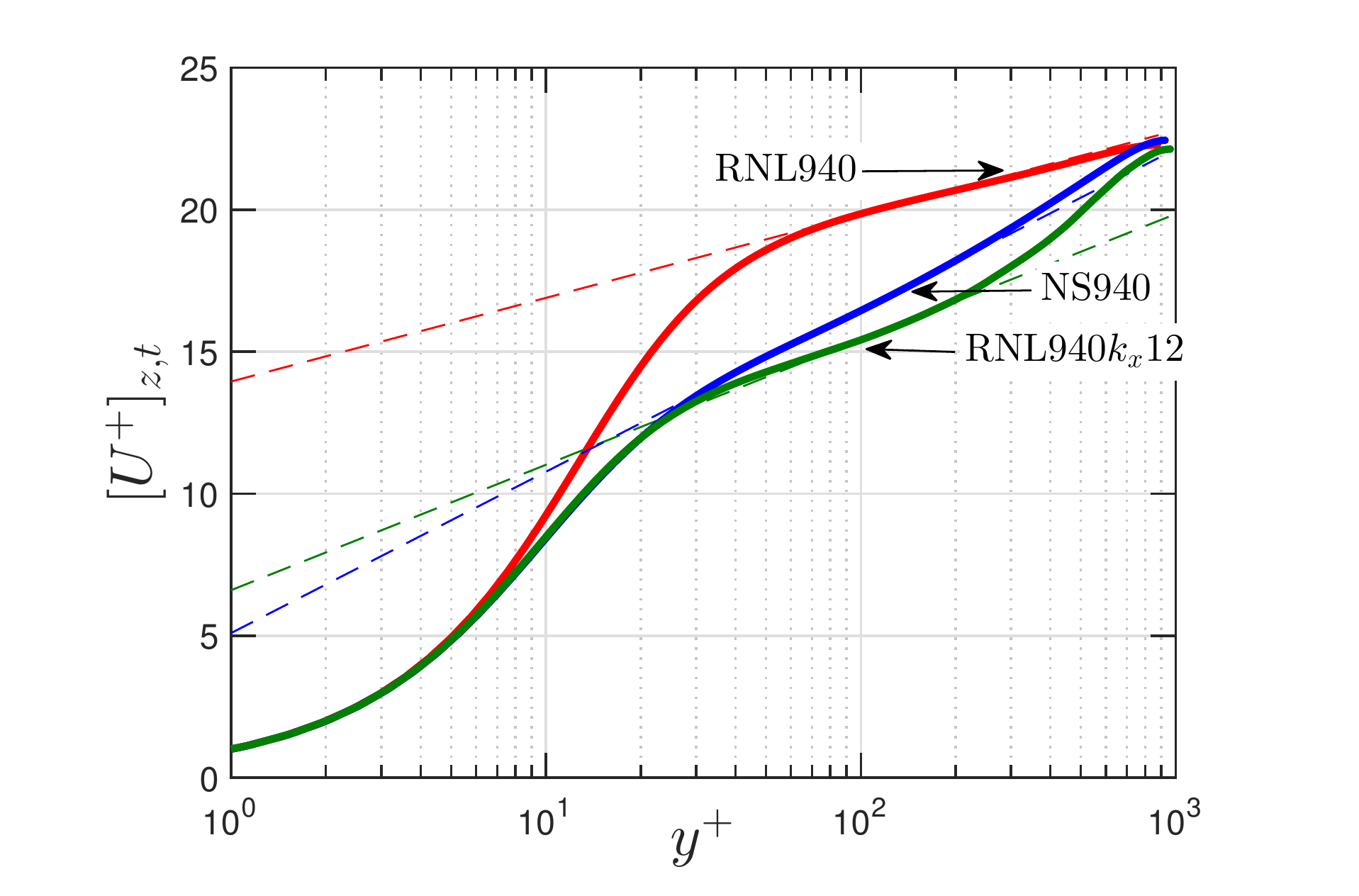}}
	\caption{\label{fig:meanU}Streamwise velocity $\[U^+\]_{z,t}$ for the simulations listed in table~\ref{table:geometry}.  The dashed lines indicate the best fit  to  the law of the wall, $\[U^+(y)\]_{z,t}=(1/\kappa) \log{(y^+)}+C$, with coefficients: $\kappa=0.40$, $C=5.1$ for NS940, $\kappa=0.77$, $C=14.0$ for RNL940 and $\kappa=0.53$, $C=6.6$ for RNL940$k_x$12.}
	\vspace{1em}
	\begin{center}
	\includegraphics[width=.7\textwidth,trim = 3mm 0mm 0mm 2mm, clip]{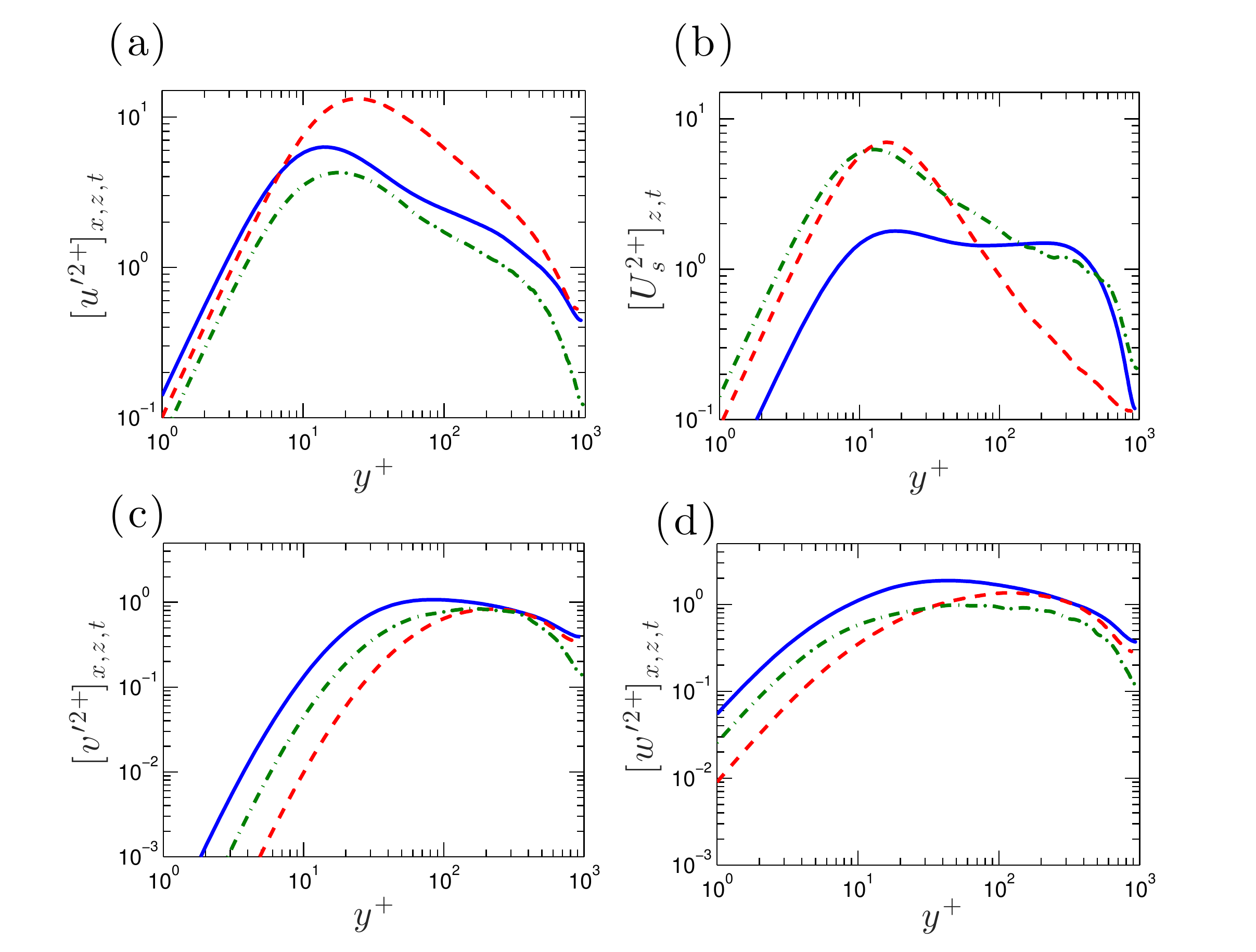}\label{fig:u}\end{center}
	\caption{Comparison of velocity fluctuations for the simulations listed in table~\ref{table:geometry}.  Shown are  (a):~$ \[ u'^{2+} \]_{x,z,t}$, (b):~$\[ U_s^{2+}\]_{z,t}$, (c):~$\[v'^{2+}\]_{x,z,t}$, (d):~$\[w'^{2+}\]_{x,z,t}$ for  NS940 (solid), RNL940 (dashed) and RNL940$k_x$12 (dash-dot).\label{fig:vel_om_prime}}
\end{figure}

\section{Comparison between NS and RNL turbulence structure and dynamics}
\label{sec:results}

In this section we compare  turbulence diagnostics obtained from
self-sustaining turbulence in the RNL system~\eqref{eq:QLE0}, to
diagnostics obtained from a parallel  associated DNS of~\eqref{eq:NSE0}
(cf. Table~\ref{table:geometry} for the parameters). The corresponding turbulent mean profiles for the NS940,  RNL940 and RNL940$k_x$12 simulations are shown in figure~\ref{fig:meanU}.

Previous simulations in Couette turbulence at lower Reynolds numbers ($\Ret=65$) showed very small difference between the mean turbulent profile in NS and RNL simulations~\citep{Thomas-etal-2014-sustain}. These simulations  at larger Reynolds numbers show significant differences in the mean turbulent profiles sustained by
NS940 and RNL940 simulations. This is especially pronounced in the outer regions where RNL940
sustains a mean turbulent profile with substantially smaller shear.

All these examples exhibit a logarithmic layer. However, the shear in these
logarithmic regions is different: the von~K\'arm\'an constant of NS at
$\Ret=940$ is $\kappa=0.4$, while for the RNL940 it is $\kappa=0.77$ and
for the  RNL940$k_x$12 it is $\kappa=0.53$. Formation of a logarithmic layer
indicates that the underlying dynamics  of the logarithmic layer are
retained in RNL. Because in the logarithmic layer RNL dynamics maintains in local balance
with dissipation essentially the same stress and variance as NS, but with
a smaller shear, RNL dynamics is in this sense more efficient  than NS in
that it produces the same local Reynolds stress while requiring less local energy input to the turbulence. To see this consider that local energy balance in the log-layer requires that  the energy production,  $ U' u_\tau^2$  (with  $U' \equiv\df\,[U]_{z,t}/\df y$) equals the energy dissipation $\epsilon$, and, because in the log-layer $U' = u_\tau/(\kappa y)$, local balance requires that $ u_\tau^3/(\kappa y) = \epsilon$, as discussed by~\citet{Townsend,Dallas-Vassilicos-2009}. This indicates that the higher $\kappa$ in RNL simulations with the same $\ut$ is associated with smaller dissipation than in the corresponding DNS. 
In RNL dynamics this local equilibrium determining the shear, and by implication $\kappa$, results from establishment of a statistical equilibrium by the feedback between the perturbation equation and the mean flow equation, with this feedback producing a $\kappa$ determined to maintain energy balance locally in $y$. These considerations imply that the $\kappa$ observed produces a local shear for which, given the turbulence structure 
produced by the restricted set of retained Fourier components in RNL, the Reynolds stress and dissipation are in local balance. Examination of the transition from  NS940 to  RNL940, shown by the simulation diagnostics in figure~\ref{fig:Enkx_DNSQL950}b, reveals the action of this feedback control associated with the reduction in  shear of the mean flow. When in~\eqref{eq:NSp} the interaction among the perturbations is switched off,  so that the simulation is governed by RNL dynamics, an adjustment occurs in which the energy of the surviving  $k_x \ne 0$ components obtain new statistical equilibrium values. An initial increase of the energy of these components is expected because the dissipative effect of the perturbation--perturbation nonlinearity that acts on these components is removed in RNL. As these modes grow, the SSP cycle adjusts to establish a new turbulent equilibrium state which is characterized by increase in energy of the largest streamwise scales and on average a reduction in  streak amplitude. In the outer layer this new equilibrium is characterized in the case of RNL940 by reduction of the shear of the mean flow and reduction in the streak amplitude (cf. figure~\ref{fig:meanU}).

A comparison of  the perturbation statistics  of RNL940  with   NS940 is shown in figure~\ref{fig:vel_om_prime}. The $u'$ component of the perturbation velocity fluctuations is significantly  more pronounced in RNL940 (cf.~figure~\ref{fig:vel_om_prime}a)
and the magnitude of the streak in RNL940 exceeds significantly the streak magnitude
in NS940 in the inner wall region (cf.~figure~\ref{fig:vel_om_prime}b). In contrast, the wall-normal and  spanwise fluctuations in RNL940 are less pronounced than in NS940 (cf.~figure~\ref{fig:vel_om_prime}c,d) and the streak fluctuations in the outer region are also less pronounced in RNL940 (cf.~figure~\ref{fig:vel_om_prime}b).

\begin{figure}
	\begin{center}
	\centerline{\includegraphics[width=4.6in,trim = 2mm 2mm 6mm 2mm, clip]{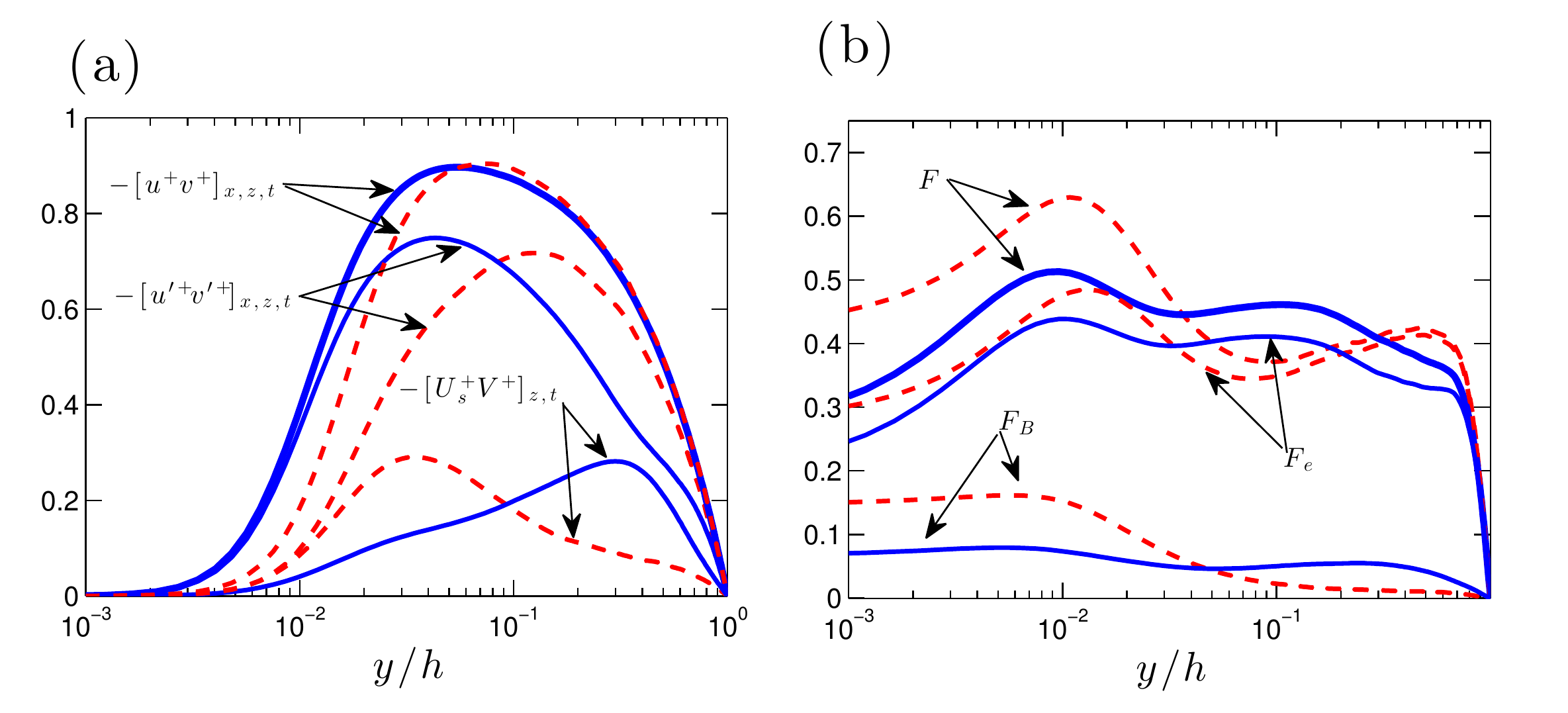}}
	\end{center}
	\caption{(a):~The Reynolds stress component, $-\[uv\]_{x,z,t}$ in NS940 (solid) and  in RNL940 (dashed). Also shown are each of the terms, $-\[ U_sV\]_{z,t} $ and $-\[u'v'\]_{x,z,t}$ that sum to $-\[uv\]_{x,z,t}$. Although the NS and RNL values of the total $-\[uv\]_{x,z,t}$ are almost identical, the contribution of  $-\[ U_sV\]_{z,t}$ and $-\[ u'v'\]_{x,z,t}$  differ in NS and RNL. (b):~Structure coefficient, $F$, in NS940 (solid) and in RNL940 (dashed). Shown are $F_B=-\[ UV \]_{z,t} \big/ \sqrt{ \[ U^2\]_{z,t} \[ V^2\]_{z,t}}$,  $F_e=-\[u'v'\]_{x,z,t} \big/ \sqrt{\[u'^2\]_{x,z,t}\[v'^2\]_{x,z,t}}$ and $F=-\[uv\]_{x,z,t} \big/ \sqrt{\[u^2\]_{x,z,t}\[v^2\]_{x,z,t}}$.\label{fig:uv_F}}
\end{figure}

\begin{figure}
	\centerline{\includegraphics[width= .56\textwidth,trim = 6mm 8mm 6mm 2mm, clip]{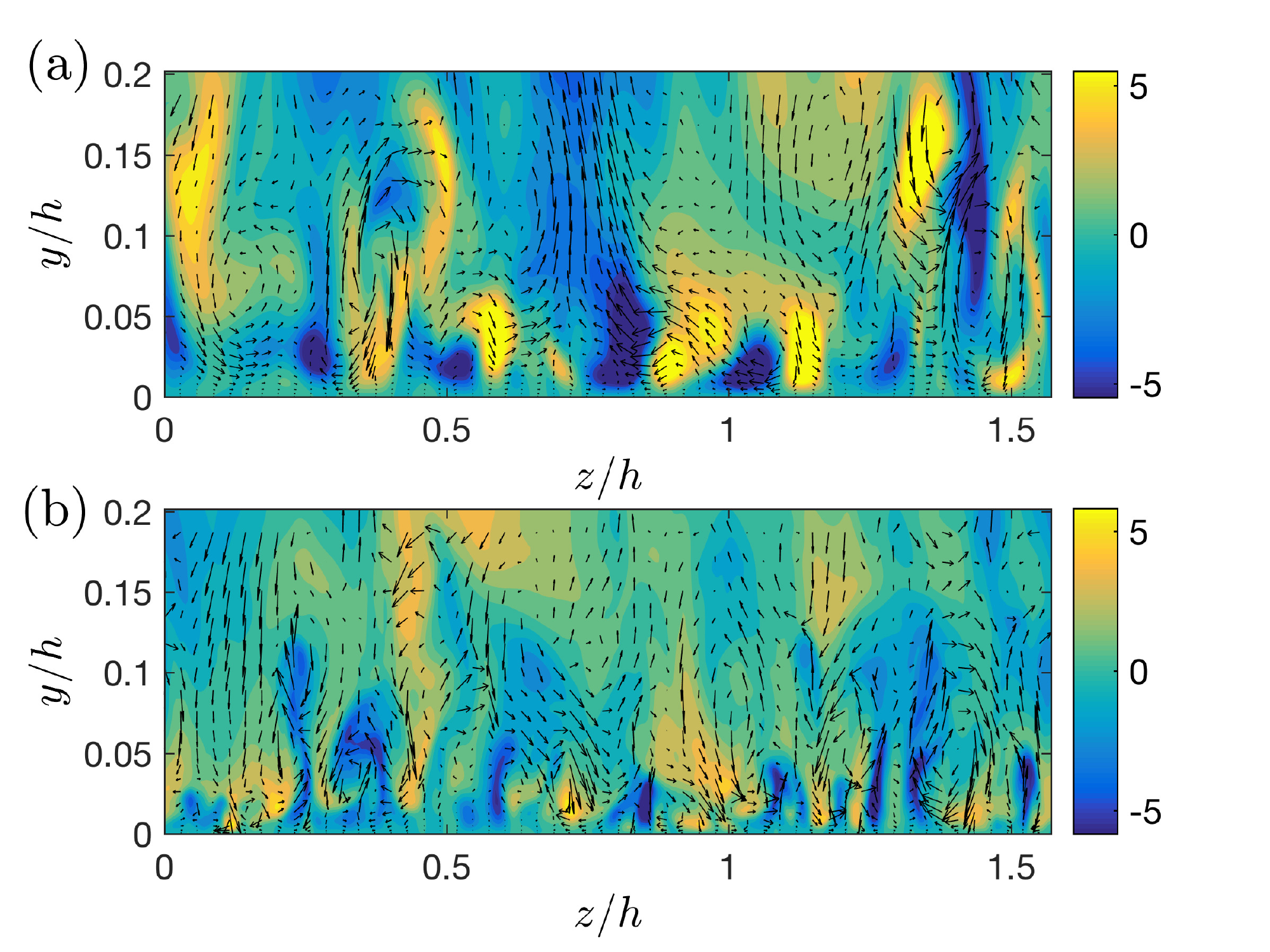}}
	\caption{Perturbation structure, $\u'^+$  in $(y,z)$ plane cross-section  for (a)~RNL940 and (b)~NS940 in the inner wall region, $0\le y/h \le0.2$. Both panels show a color mapping
	 of the $u'^+$ field, superimposed with $(v'^+,w'^+)$ velocity vectors.\label{fig:YZpert}}
	\vspace{.5cm}
	\centerline{\includegraphics[width= .98\textwidth,trim = 10mm 1mm 10mm 0mm, clip]{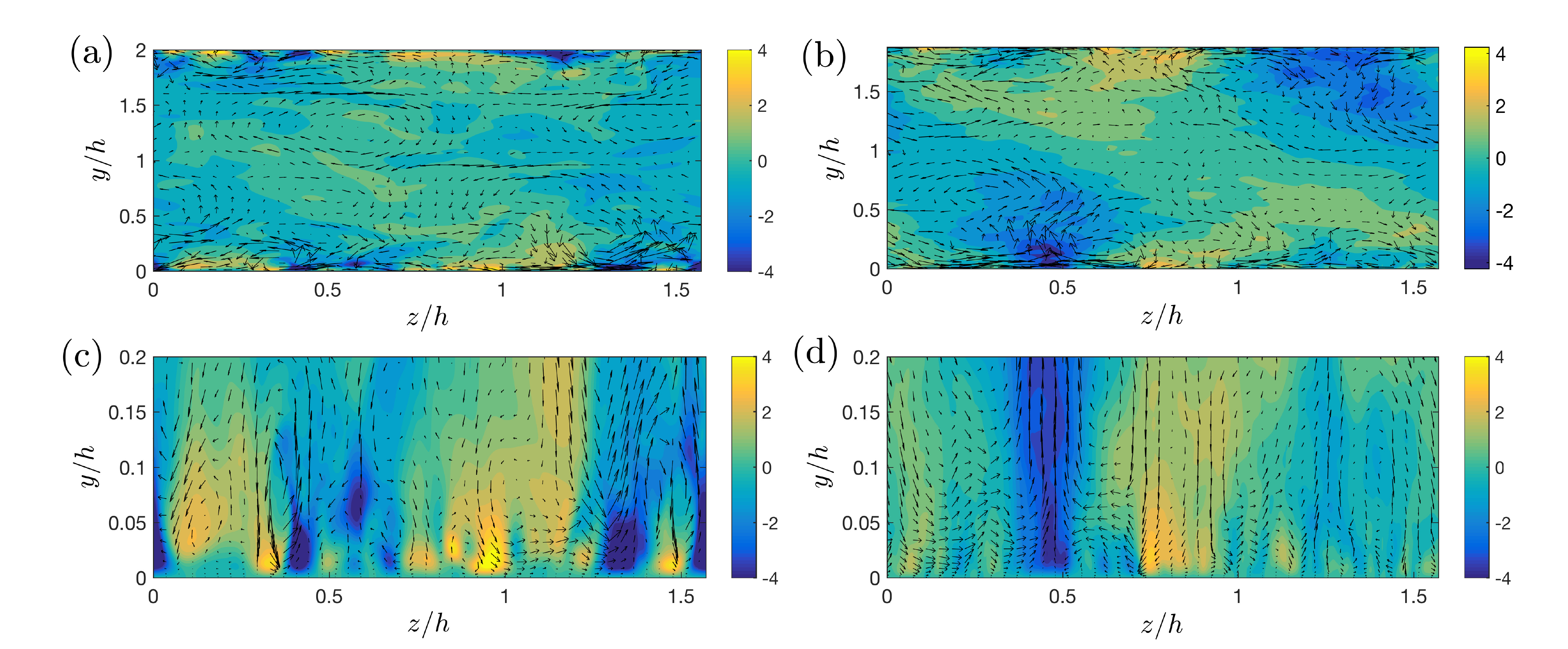}}
	\caption{Instantaneous streak component of the flow, $U_s^+$, shown as a $(y,z)$ plane cross-section for (a), (c) RNL940 and (b), (d) NS940. All panels show a color mapping of the streak velocity, $U_s^+$, superimposed with  $(V^+,W^+)$ velocity vectors. Panels (a) and (b)  show the whole channel while panels (c) and (d) show the inner wall region, $0\le y/h\le 0.2$.\label{fig:YZstreak}}
\end{figure}


\begin{figure}
  \centering
  \subfloat{\label{fig:dns_perspective}}{
   \includegraphics[width = .9\textwidth,trim = 3mm 7mm 3mm 8mm, clip]{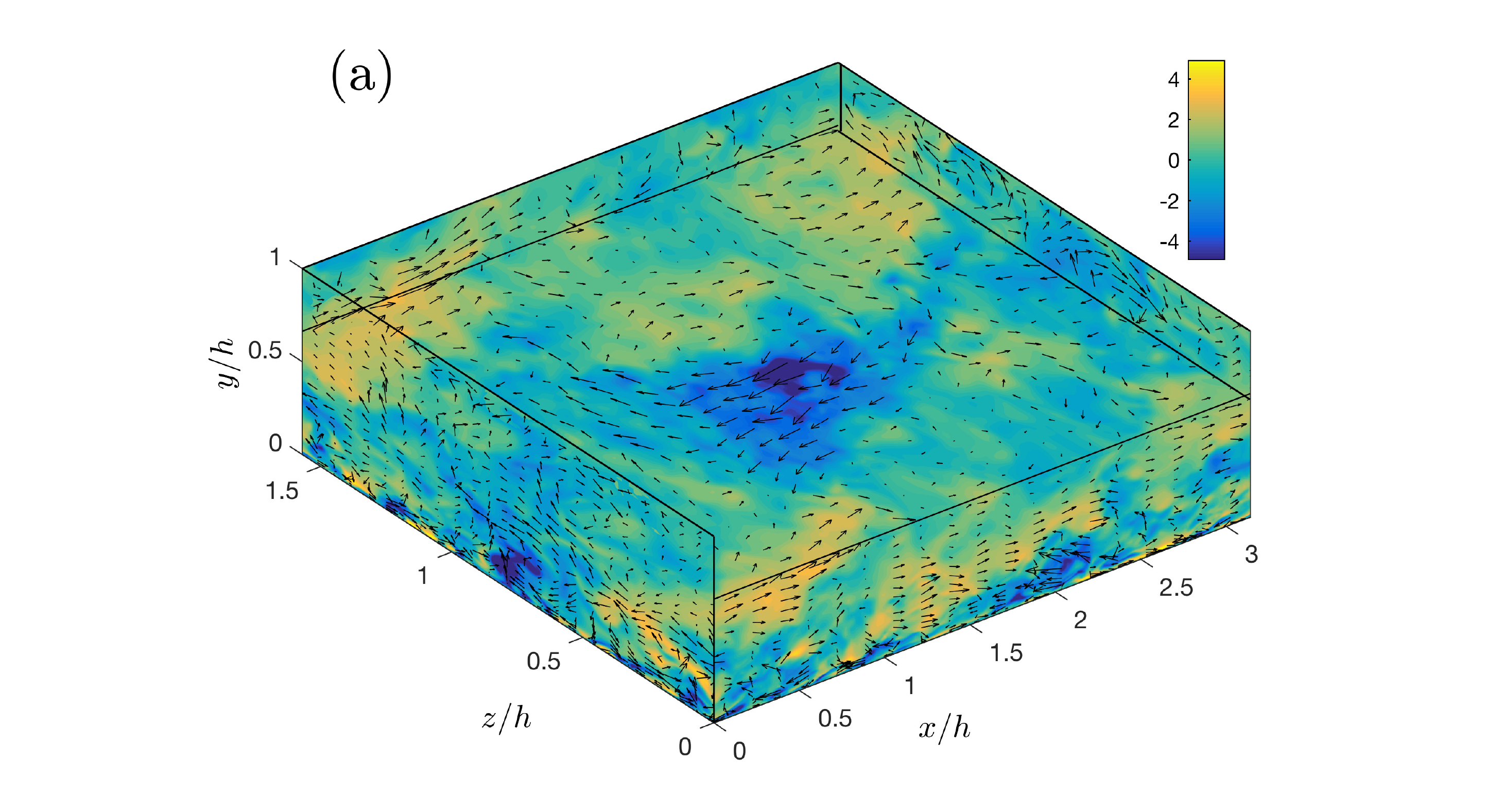}}
  \subfloat{\label{fig:QL_perspective}}{
   \includegraphics[width = .9\textwidth,trim = 3mm 7mm 3mm 8mm, clip]{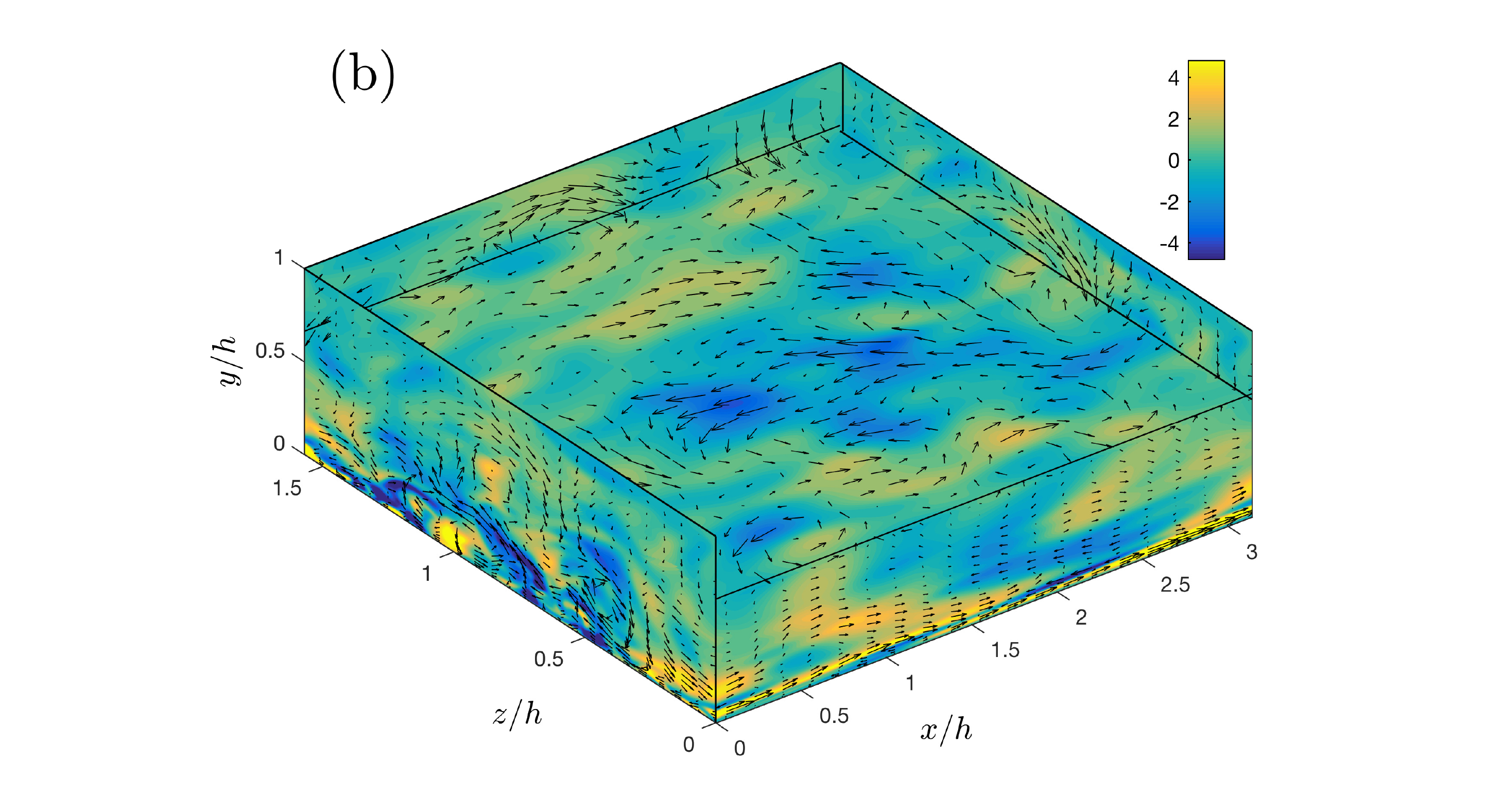}}
	 \caption{3D perspective plots of the flow  at a single time for (a)~NS940, and (b)~RNL940 for lower half of the channel, $0\le y/h \le 1$. Both images show a color mapping of the streak component plus streamwise perturbation, $U_s^++u'^+$. The central $x$-$z$ panel shows the flow at channel height, $y/h =0.65$. The superimposed vectors represent the $(U_s^++u'^+,w^+)$ velocities for the $(x,z)$-plane, $(U_s^++u'^+,v^+)$ velocities for the $(x,y)$-plane and $(v^+,w^+)$ velocities for the $(y,z)$-plane. The parameters of the simulations are given in table~\ref{table:geometry}.\label{fig:3Dstreak}}
\end{figure}

Despite these differences in the r.m.s. values of the velocity fluctuations,  both RNL940 and NS940 produce very similar Reynolds stress  $-\[ uv \]_{x,z,t}$, which is the sum  of $-\[U_s V\]_{z,t}$ and  $-\[ u'v' \]_{x,z,t}$.
Comparison of the wall-normal distribution  of these two components of the Reynolds stress is shown in figure~\ref{fig:uv_F}a. Because the turbulence in NS940  and  RNL940 is sustained with essentially the same pressure gradient, the sum of these Reynolds stresses  is the same linear function of $y$ outside the viscous layer.  The Reynolds stress is dominated by the perturbation Reynolds stress $-\[ u'v' \]_{x,z,t}$, with the RNL stress penetrating farther from the wall. This is consistent with the fact that the perturbation structure in RNL has larger scale. This can be seen in a comparison of the NS and RNL perturbation structure  shown in figure~\ref{fig:YZpert}. Note that the Reynolds stress $-\[ U_sV\]_{z,t}$ associated with the streak and roll in the outer region of the NS940 simulation is larger than that in RNL940. Further, the average correlation between the perturbation $u'$ and $v'$ fields are  almost the same in both simulations, while the correlation between the $U_s$ and $V$ in RNL940 is much smaller than that in NS940 in the outer layer. This is seen in a plot of the structure coefficient (cf.~\citet{Flores-Jimenez-2006}) shown  in figure~\ref{fig:uv_F}b.

Turning now to the flow structures in the NS940 and RNL940 simulations, a $(y,z)$-plane  snapshot of the streamwise mean flow component (corresponding to $k_x=0$ streamwise wavenumber) is shown in figure~\ref{fig:YZstreak}. A color mapping
of the streamwise streak component, $U_s$, is shown 
together with vectors of the  streamwise mean $(V,W)$ field, which indicates the velocity components of the large-scale  roll structure. The presence of  organized streaks and associated rolls is evident both in the inner-wall and in the outer-wall region.  Note that, in comparison with the streak in NS940,  the streak in RNL940 has a finer $(y,z)$
structure, which is consistent with  the energy of the streak being more strongly dissipated by diffusion in RNL (cf. figure~\ref{fig:YZstreak}). A three-dimensional perspective of the flow in  NS940 and RNL940 is shown in figure~\ref{fig:3Dstreak}. Note that in RNL940  there is no visual evidence of the $k_x=0$ roll/streak structure which is required by the restriction of RNL dynamics to be  the primary structure responsible for organizing and maintaining the self-sustained turbulent state.  Rather, the most energetic structure among the perturbations maintaining the pivotal streamwise mean roll/streak is the structure that dominates the observed turbulent state. We interpret this as indicating that the $k_x=0$ roll/streak structure, which is the dynamically central organizing structure in RNL turbulence and which organizes the turbulence on scale unbounded in the streamwise direction, cannot be reliably identified by visual inspection of the flow fields, which would lead one to conclude that the organizing scale was not just finite but the rather short scale of the separation between perturbations to the streak. Essentially this same argument is cast in terms of the inability of Fourier analysis to identify the organization scale of the roll/streak structure by~\citet{Hutchins-Marusic-2007}. This dynamically central structure,  which appears necessarily at $k_x=0$ in
RNL dynamics,  is reflected in the highly streamwise elongated structures seen in simulations and
observations of DNS wall turbulence. While in a long channel averaging would be expected to suppress the
$k_x=0$ component in DNS, in the short channel used here,  the $k_x=0$ component is prominent in both the RNL940 and NS940 (cf. figure~\ref{fig:YZstreak}).

An alternative view of turbulence structure is provided by comparison of the spectral energy densities of velocity fields as a function of streamwise and spanwise wavenumber, $(k_x,k_z)$.  The premultiplied spectral energy densities of each of the three components of velocity, $E_{uu}$, $E_{vv}$ and $E_{ww}$, are shown at heights $y^+=20$, representative of the inner-wall region; and at $y/h =0.65$, representative of the outer-wall region, in figure~\ref{fig:spectr_2d}. While RNL940 produces spanwise streak spacing and rolls similar to those in NS940, the tendency of RNL to produce longer structures in this diagnostic is also evident. The spectra for the outer region indicate similar large-scale structure and good agreement in the spanwise spacing between RNL940 and NS940. This figure establishes the presence of large-scale structure in the outer region in both RNL940 and NS940.
It has been noted that in NS940, while the scale of the structures increases
linearly with distance from the wall in the inner-wall region, in the outer regions the structures having the largest possible streamwise scale dominate the flow variance at high Reynolds number~\citep{Jimenez-1998,Jimenez-Hoyas-2008}. This linear scaling near the wall can also be seen in figure~\ref{fig:lambda_z}, where contour plots of normalized premultiplied one-dimensional spectral energy densities as a function of spanwise wavelength, $k_z$, and wall-normal distance, as in~\citet{Jimenez-1998,Jimenez-Hoyas-2008}, are shown for
NS940 and RNL940. In both simulations, the spanwise wavelength associated with the spectral density maxima increases linearly with wall distance, with this linear dependence being interrupted
 at $y/h\approx 0.5$ (or $y^+\approx 450$). Beyond $y/h\approx 0.5$ , structures assume the 
 largest $\lambda_z$   allowed in the channel, suggesting simulations be performed in larger boxes in future work (cf.~discussion by~\citet{Jimenez-Hoyas-2008} and~\citet{Flores-Jimenez-2010}). Corresponding contour plots of spectral energy density as a function of streamwise wavelength and wall-normal distance  are shown in figure~\ref{fig:lambda_x}. These plots show that the perturbation variance in the inner wall and outer wall region is concentrated in a limited set of  streamwise components. In the case of RNL940 the  spontaneous restriction on streamwise perturbation wavenumber support that occurs in RNL dynamics produces a corresponding sharp shortwave cutoff in the $k_x$ components of the  spectra, as seen in panels (d,e,f) of figure~\ref{fig:lambda_x}. 
Note that the maximum wavelength in these graphs is equal to the streamwise length of the box, and not to the infinite wavelength associated with the energy of the roll/streak structure in RNL dynamics.
%

\begin{figure}
\centerline{\includegraphics[width=5in,trim = 0mm 20mm 0mm 0mm, clip]{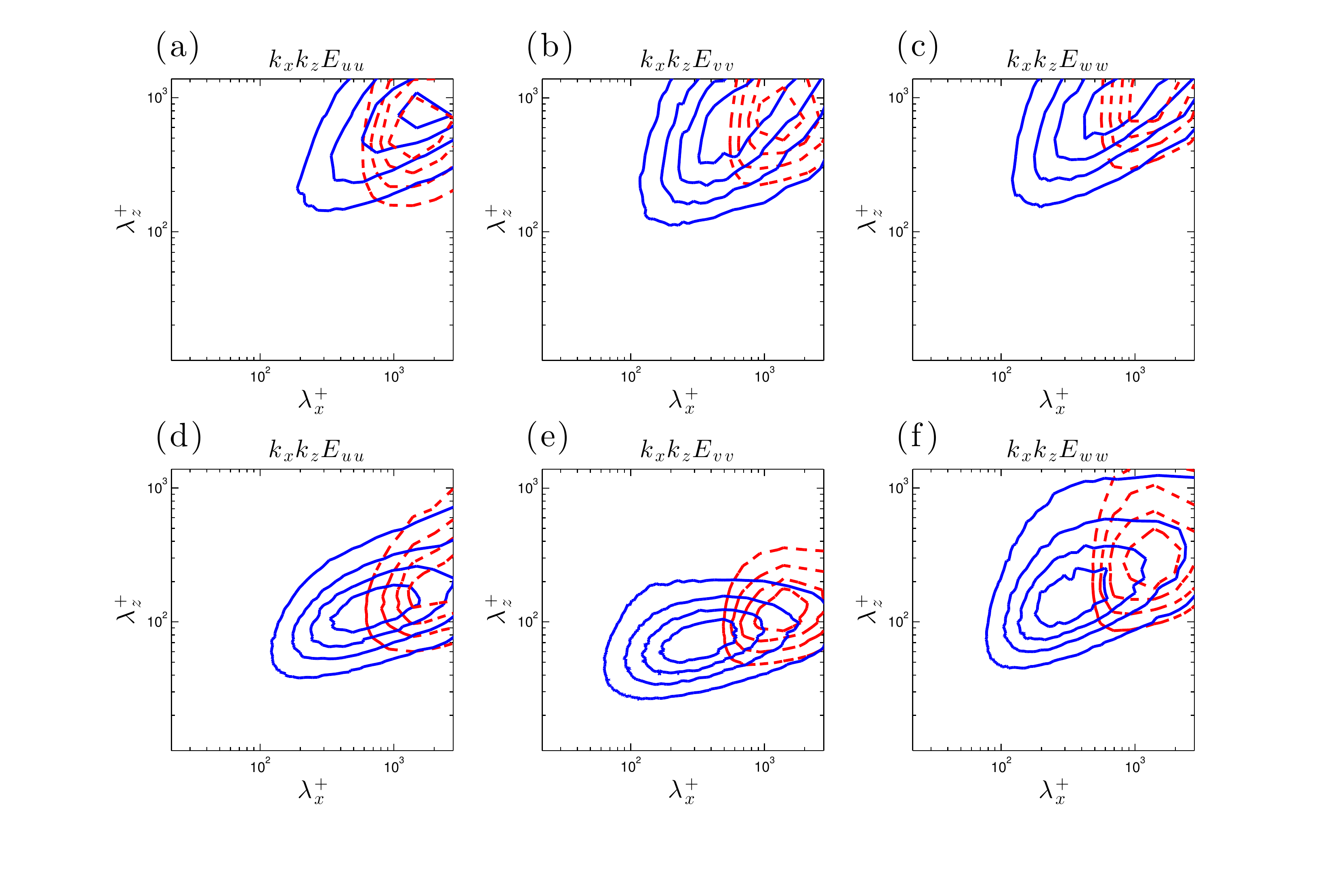}}
\caption{Contours of pre-multiplied power spectra $k_x k_z E_{ff}(k_x,k_z)$ with $f=u,v,w$, as a function of $\la^+_x$ and $\la^+_z$ for NS940 (solid) and RNL940 (dashed). Panels (a), (b) and (c) show the spectral energy densities at wall distance $y/h=0.65$ for the $u$, $v$ and $w$ respectively, while panels (d), (e) and (f) show the corresponding spectral energy densities at $y^+=20$. Contours are (0.2,0.4,0.6,0.8) times the maximum value of the corresponding spectrum. The maximum $\la_x^+$ and $\la_y^+$ are the lengths $L_x^+$, $L_z^+$ of the periodic channel.\label{fig:spectr_2d}}
\end{figure}
%

\begin{figure}
\begin{center}
\includegraphics[width=.9\textwidth,trim = 45mm 2mm 15mm 2mm, clip]{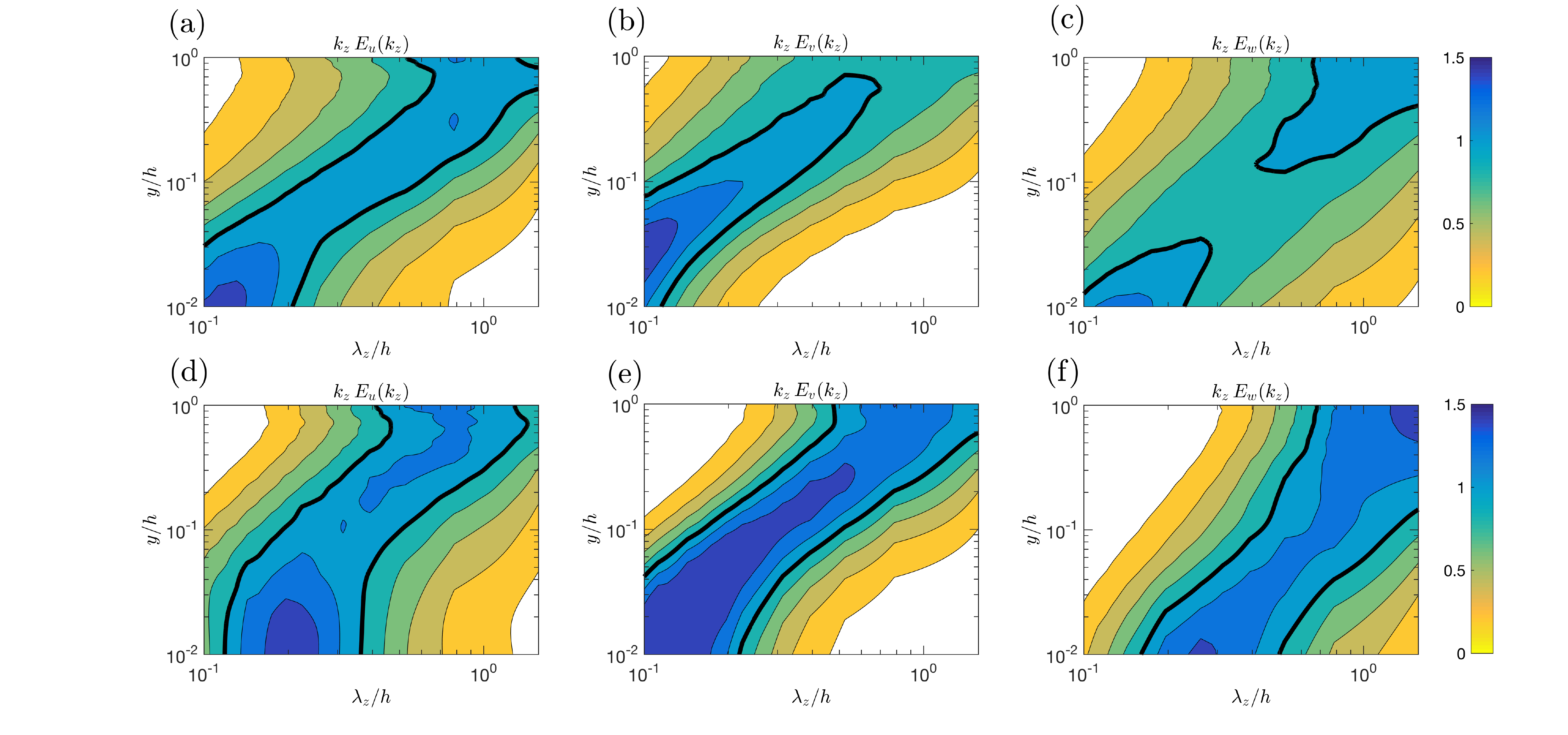}
\end{center}
\caption{Normalized pre-multiplied spectral densities $k_z E_{f}(k_z) = k_z \sum_{k_x} E_{ff}(k_x,k_z)$, with $f=u,v,w$, as a function of spanwise wavelength, $\la_z/h$, and $y/h$. Spectral densities are normalized so that at each $y$ the total energy, $\sum_{k_z} E_{f}(k_z)$, is the same.  Shown are for NS940 (a):~$k_z E_{u}(k_z)$, (b):~$k_z E_{v}(k_z)$, (c):~$k_z E_{w}(k_z)$ and for RNL940 (d):~$k_z E_{u}(k_z)$, (e):~$k_z E_{v}(k_z)$, (f):~$k_z E_{w}(k_z)$. The isocontours are $0.2,0.4,\dots,1.4$ and the thick line marks the 1.0 isocontour.\label{fig:lambda_z}}
\vspace{1.5em}
\begin{center}
\includegraphics[width=.9\textwidth,trim = 45mm 2mm 15mm 2mm, clip]{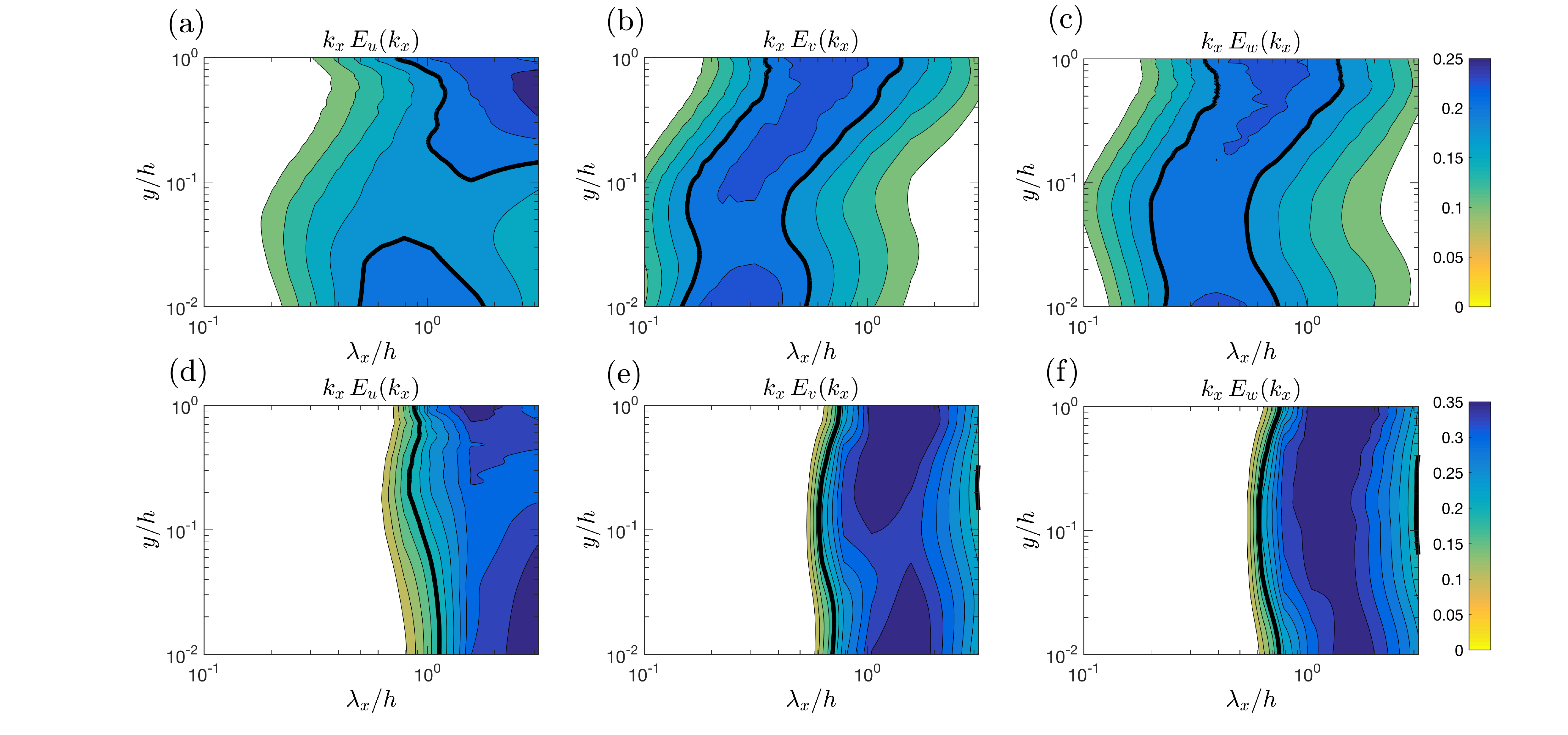}
\end{center}
\caption{Normalized pre-multiplied spectral densities $k_x E_{f}(k_x) = k_x \sum_{k_z} E_{ff}(k_x,k_z)$, with $f=u,v,w$, as a function of streamwise wavelength, $\la_x/h$, and $y/h$. Spectral densities are normalized so that at each $y$ the total energy, $\sum_{k_x} E_{f}(k_x)$, is the same. Shown are for NS940 (a):~$k_x E_{u}(k_x)$, (b):~$k_x E_{v}(k_x)$, (c):~$k_x E_{w}(k_x)$ and for RNL940 (d):~$k_x E_{u}(k_x)$, (e):~$k_x E_{v}(k_x)$, (f):~$k_x E_{w}(k_x)$. The isocontours are $0.1,0.125,\dots,0.35$ and the thick line marks the 0.2 isocontour.\label{fig:lambda_x}}
\end{figure}

\section{Streak structure dynamics in NS and RNL dynamics}

That RNL dynamics maintains a turbulent state similar to that of NS with nearly the same $\Ret$  ($\Ret=882$  with the six Fourier components of RNL940 and $\Ret=970.2$ for the single Fourier component with $k_x h =12$  of RNL940$k_x$12 versus the $\Ret=940$ of the NS940; cf.~table~\ref{table:geometry}) implies that these systems have approximately the same energy production and dissipation and that the reduced set of Fourier components retained in RNL dynamics assume the burden of accounting for this energy production and dissipation. Specifically, the components in NS940 that are not retained in RNL dynamics are responsible for approximately one-third of the total energy dissipation, which implies that the components that are retained in RNL940 dynamics must increase their dissipation, and consistently their amplitude, by that much.

Large scale roll/streak structures are prominent  in the inner layer as well as in the outer layer both in NS940 and in RNL940. In the inner layer, the interaction of roll/streak structures with the $k_x\ne0$ perturbation field  maintains turbulence through an SSP~\citep{Hamilton-etal-1995, Jimenez-Pinelli-1999,Farrell-Ioannou-2012}. The RNL system provides an especially simple manifestation of this SSP, as its dynamics comprise only interaction between the mean ($k_x=0$) and perturbation ($k_x \ne 0$) components. The fact that RNL self-sustains a close counterpart of the NS turbulent state in the inner wall region provides strong evidence that the RNL SSP captures the essential dynamics of turbulence in this region.

The structure of the RNL system compels the interpretation that the time dependence of the SSP cycle in this 
system seen in figure \ref{fig:Enkx_DNSQL950}(b) is an intricate interaction of dynamics among streaks, rolls and perturbations that produces the time dependent streamwise mean flow $\U(y,z,t)$, 
which, when introduced in~\eqref{eq:QLpf}, results in generation of a particular evolving perturbation Lyapunov structure with exactly zero Lyapunov exponent
that simultaneously produces Reynolds stresses contrived to maintain the associated  time dependent mean flow.
S3T identifies this exquisitely contrived SSP cycle  comprising the generation of the
streak  through lift-up by the rolls, the maintenance of the rolls by torques
induced by the perturbations which themselves are maintained by time-dependent parametric
non-normal interaction with the streak~\citep{Farrell-Ioannou-2012}. 


In RNL this SSP  is more efficient than its DNS counterpart in producing downgradient perturbation momentum 
flux, as with smaller mean shear over most of the channel,  a self-sustained turbulence with approximately the same $\Ret$ as that in NS940 is maintained, as discussed above (cf.~section~\ref{sec:results}).
A comparison of the shear, the r.m.s. $V$ velocity, and the r.m.s. streak velocity, $U_s$, in the outer layer is shown as a function of $y$ in figure~\ref{fig:UyVrmsUsrms}, from which it can be seen that the ratio of the product of the mean shear and r.m.s. wall normal velocity to the r.m.s. streak velocity is approximately equal in DNS and in RNL. It is important to note that in the RNL system these dependencies arise due to  the feedback control exerted by the perturbation dynamics on the mean flow dynamics by which its statistical steady state is determined. The structure of RNL isolates this feedback control process so that it can be studied, and elucidating its mechanism and properties are the subject of ongoing work.

\begin{figure}
	\begin{center}
	\includegraphics[width=.9\textwidth]{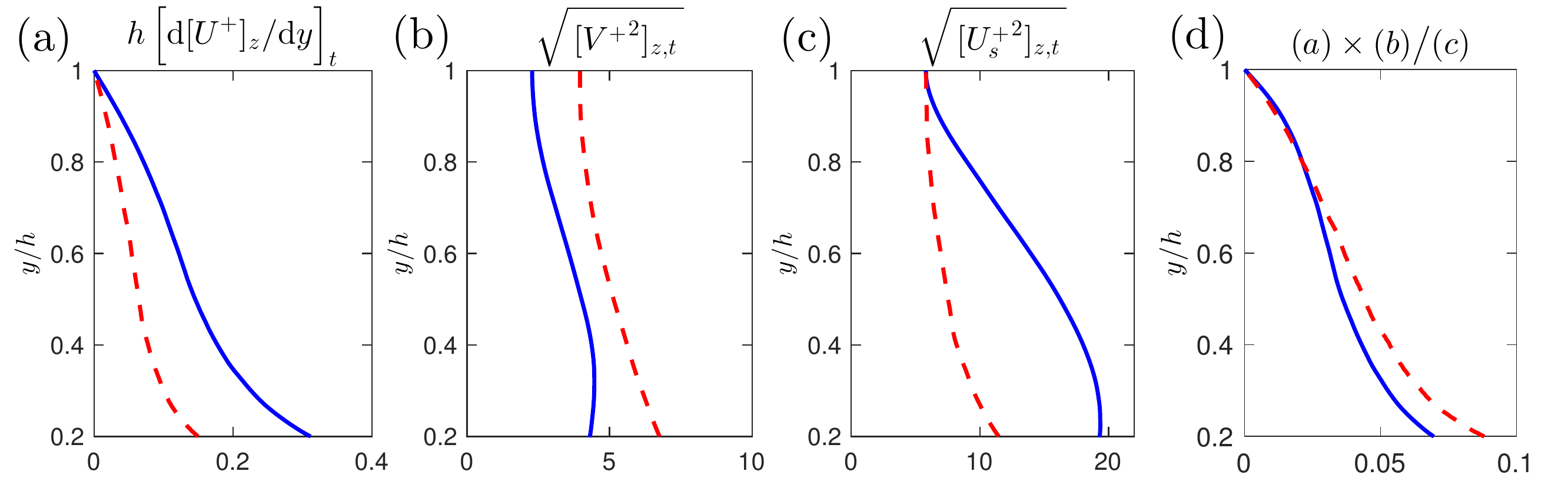}
	\end{center}
	\caption{Comparison of the (a) turbulent mean shear, $\[\df [ U^+]_z/\df y\]_t h$, (b) the r.m.s.~of $\[V^+\]_z$, (c) the r.m.s.~of the streak velocity, $U_s^+$ and  (d) the ratio of the product of the mean shear and r.m.s. wall normal velocity over the r.m.s. streak velocity, for  NS940 (solid) and RNL940 (dashed) in the outer layer, $0.2\le y/h\le 1$.\label{fig:UyVrmsUsrms}}
\end{figure}

In the discussion above we have assumed that the presence of roll and streak structure in the log-layer in RNL indicates the existence of an SSP cycle there, and by implication also in NS.  In order to examine this SSP,  consider the momentum equation for the streamwise streak:
\begin{align}
	\partial_t U_{s} &=\underbrace{ -\(V \,\partial_y U -\[V \,\partial_yU\]_z\bit\) -\( W \,\partial_z U-\[ W \,\partial_z U\]_z\bit\)}_{\textrm{A}} \nonumber\\
	&\qquad\qquad \underbrace{-\(\[ v'  \,\partial_y u'-\[v'  \,\partial_y u'\]_z\bit\]_x\)- \(\[ w'  \,\partial_z u'-\[w'  \,\partial_z u'\]_z \bit\]_x\)}_{\textrm{B}}+ \underbrace{\vphantom{-\(V \,\partial_y U -\[V \,\partial_yU\]_z\bit\) }\nu\, \Del U_{ s}}_{\textrm{C}}\ .\label{eq:Us}
\end{align}
Term~A in~\eqref{eq:Us} is the contribution to the streak acceleration  by the `lift-up' mechanism and the `push-over' mechanism,  which represent transfer to streak momentum  by  the mean wall-normal and spanwise velocities, respectively; Term~B in~\eqref{eq:Us} is the contribution to the streak momentum by the perturbation  Reynolds stress  divergence (structures with $k_x \ne 0$); Term~C is the diffusion of the streak momentum due to viscosity.

\begin{figure}
	\centering
	\includegraphics[width= \textwidth]{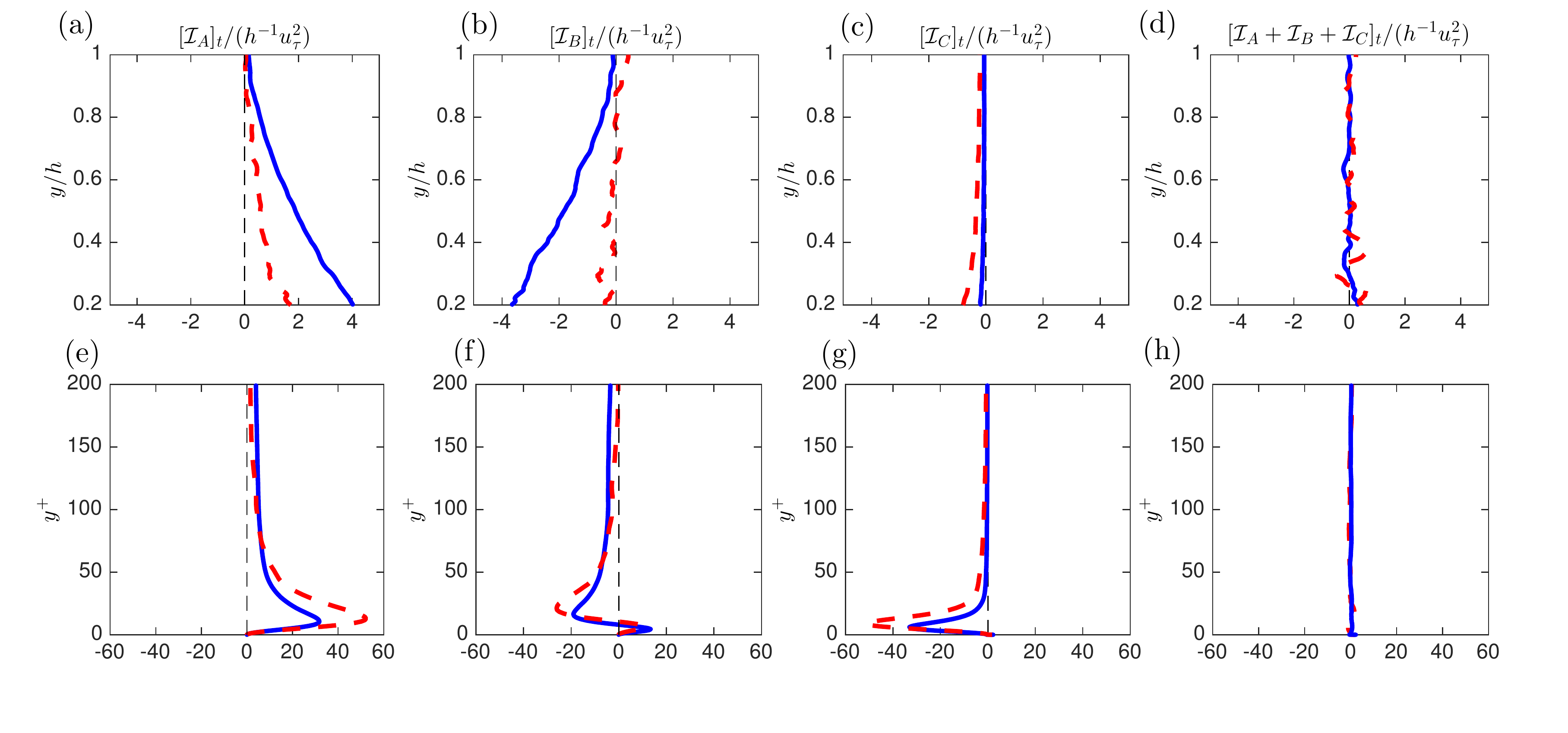}
	\vspace{-2em}
	\caption{Cross-stream structure of the time averaged contributions to streak acceleration for NS940 (solid) and RNL940 (dashed) from: (a,e) the lift-up mechanism $[\mathcal{I}_A]_t (y)$, (b,f) the perturbation Reynolds stress divergence  $[ \mathcal{I}_B]_t(y)$ and (c,g) the momentum diffusion $[\mathcal{I}_C]_t(y)$. Upper panels show structure in the outer layer, $0.2\le y/h\le 1$, lower panels show the structure in the inner layer, $0\le y^+\le 200$. In (d,h) we plot the sum of these terms which averaged over a long time interval should add exactly to zero. \label{fig:UsdotyOMxdoty}}
\end{figure}

\begin{figure}
	\centering
	\includegraphics[width= .85\textwidth,trim = 17mm 5mm 22mm 2mm, clip]{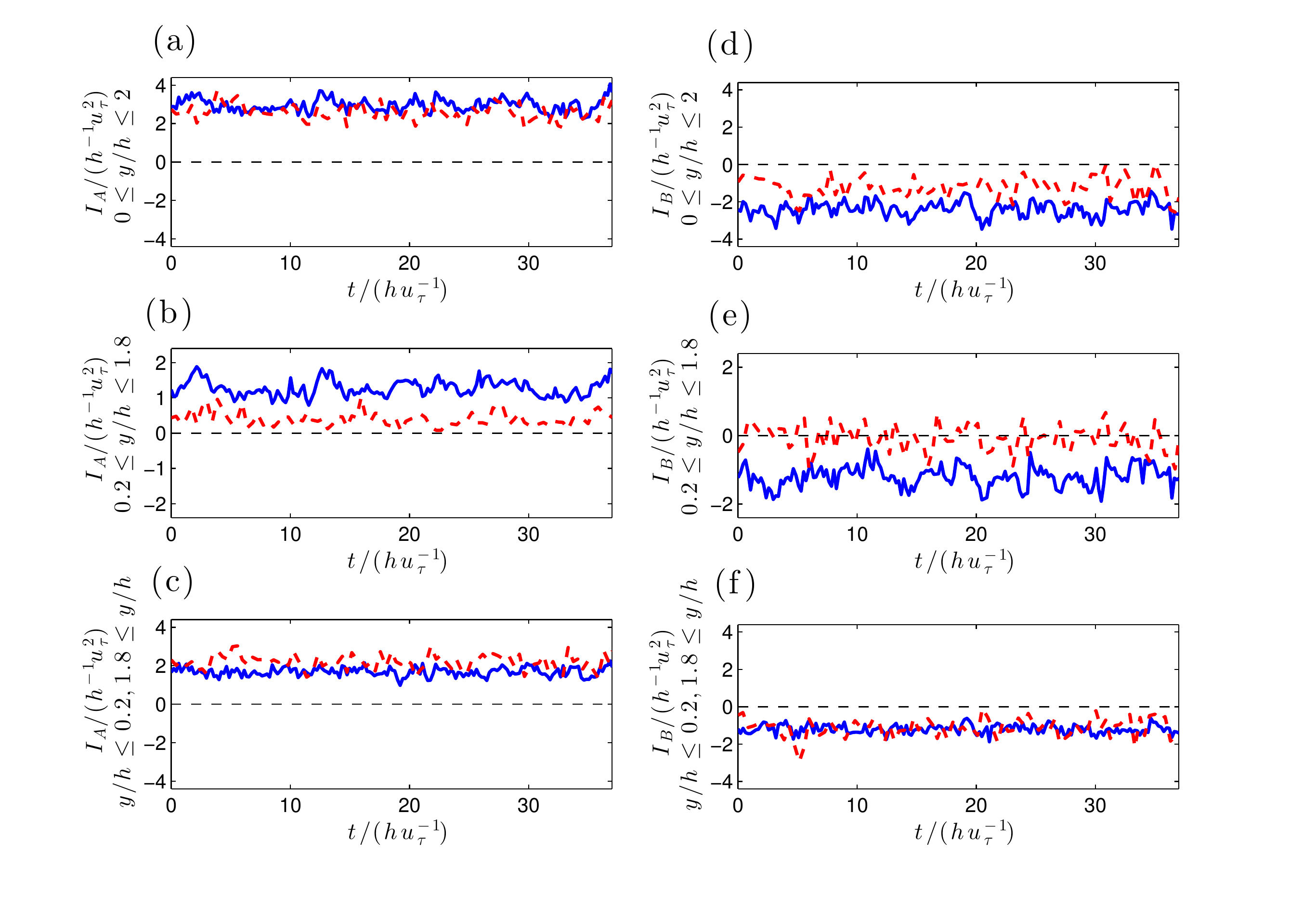}
	\caption{Time series of  the total $I_A(t)$ 
	 (lift-up) for NS940 (solid) and RNL940 (dashed),  (a): over the whole channel,
	(b): over the outer region, $0.2\le y/h\le 1.8$, (c): over the inner region, $0\le y/h\le 0.2$ and $1.8\le y/h\le 2$. Similarly for $I_B(t)$  
	 (Reynolds stress divergence), (d): over the whole channel, (e): over the outer region, $0.2\le y/h\le 1.8$, (f): over the inner region, $0\le y/h\le 0.2$ and $1.8\le y/h\le 2$.\label{fig:Usdot_time}}
\end{figure}


In order to identify the mechanism of streak maintenance we determine the contribution of
terms~A,~B and~C in~\eqref{eq:Us} to the streak momentum budget by evaluating these contributions.
The time-averaged results are shown as a function of $y$ over these
cross-stream regions of the flow, indicated by $R$: the whole channel, 
the outer region, $0.2 \le y/h \le 1.8$,  and the inner region, $0\le y/h<0.2$ and $1.8<y/h\le 2$,  in figure~\ref{fig:UsdotyOMxdoty}. The contributions are, respectively, the lift up:
\begin{equation}
	I_A (t) = h^{-1}\int_R\df y\;\mathcal{I}_{A}(y,t)\ ,\ \ {\rm with}\ \ \mathcal{I}_{A}(y,t) =\[ \bit\sgn(U_s) \times \textrm{(Term~A)}\]_z\ ,
\end{equation}
the perturbation Reynolds stress divergence:
\begin{equation}
 I_B (t) = h^{-1}\int_R\df y\;\mathcal{I}_{B}(y,t)\ ,\ \ {\rm with}\ \ \mathcal{I}_{B}(y,t) =\[ \bit\sgn(U_s) \times \textrm{(Term~B)}\]_z\ ,
\end{equation}
and diffusion:
\begin{equation}
 I_C (t) = h^{-1}\int_R\df y\;\mathcal{I}_{C}(y,t)\ ,\ \ {\rm with}\ \ \mathcal{I}_{C}(y,t) =\[ \bit\sgn(U_s) \times \textrm{(Term~C)}\]_z\ .
\end{equation}
In the inner-wall and outer-wall regions, in both NS940 and RNL940, the streak is 
maintained only by the lift-up mechanism, while streak momentum is lost on average at all cross-stream levels to both the Reynolds stress divergence and the momentum diffusion. In RNL940 the magnitude 
of streak acceleration by lift-up is greater than that of NS940 in the inner region, whereas  
in the outer region the 
acceleration by lift-up in RNL940 is about half that in  NS940,  consistent with their similar 
roll amplitude (cf. figure~\ref{fig:UyVrmsUsrms}b) and the smaller mean flow 
shear maintained at statistical steady state in RNL940. In the outer region of the 
NS940 the Reynolds stress divergence  almost completely balances the positive contribution 
from lift-up, while in RNL940 the lift up is balanced equally by the Reynolds stress 
divergence and the diffusion.
Enhancement of the contribution by diffusion in the outer layer in RNL940 results from 
the increase in the  spanwise and cross-stream wavenumbers of the streak  (cf. figure~\ref{fig:YZstreak}c) resulting from the nonlinear advection of the streak by the $V$ and $W$ velocities.
This increase in the spanwise and cross-stream wavenumbers of the streak in RNL940 due to nonlinear advection by the mean $(V,W)$ roll circulation also implies that the dissipation of  streak energy  in RNL940 is similarly enhanced. This constitutes an alternative route for energy transfer to the dissipation scale, which  continues to be available for establishment of statistical equilibrium in RNL940 despite  the limitation in the streamwise wavenumber support inherent in RNL turbulence. The lift-up process is a positive contribution 
to the maintenance of the streak, and the Reynolds stress divergence 
is a negative contribution  not 
only in a time-averaged sense, but also at every time instant. This is shown 
in plots of  the time series of  the lift up and Reynolds stress divergence 
contribution to the streak momentum over the inner region $0\le y/h<0.2$ and $1.8<y/h\le2$, 
over the outer region $0.2 \le y/h \le 1.8$, and  over  the  whole channel in figure~\ref{fig:Usdot_time}.
We conclude that, in both NS940 and RNL940,  the sole positive contribution to the outer layer streaks is lift-up, despite the small shear in this region. Consistently, a recent POD analysis in a similar flow setting has confirmed the phase relationship between the streak and wall-normal velocity, indicative of this lift-up mechanism \citep{Nikolaidis-etal-Madrid-2016}. We next consider the dynamics maintaining the lift-up.

\section{Roll dynamics: maintenance of mean streamwise vorticity in NS and RNL\label{sec:roll}}

We have established that the lift-up mechanism is not only responsible for streak maintenance in the 
inner layer, but also in the outer layer. We now examine the mechanism of  the lift-up by relating it to maintenance of the roll structure using as a diagnostic streamwise-averaged vorticity, $\Omega_x=\partial_yW-\partial_zV$. In order for roll circulation to be maintained against dissipation there must be a continuous generation of $\Omega_x$. There are two possibilities for the maintenance of $\Omega_x$ in the outer layer: either $\Omega_x$ is generated locally in the outer layer, or it is advected  from the near-wall region.

\begin{figure}
\centering
\centerline{\includegraphics[width= .9\textwidth]{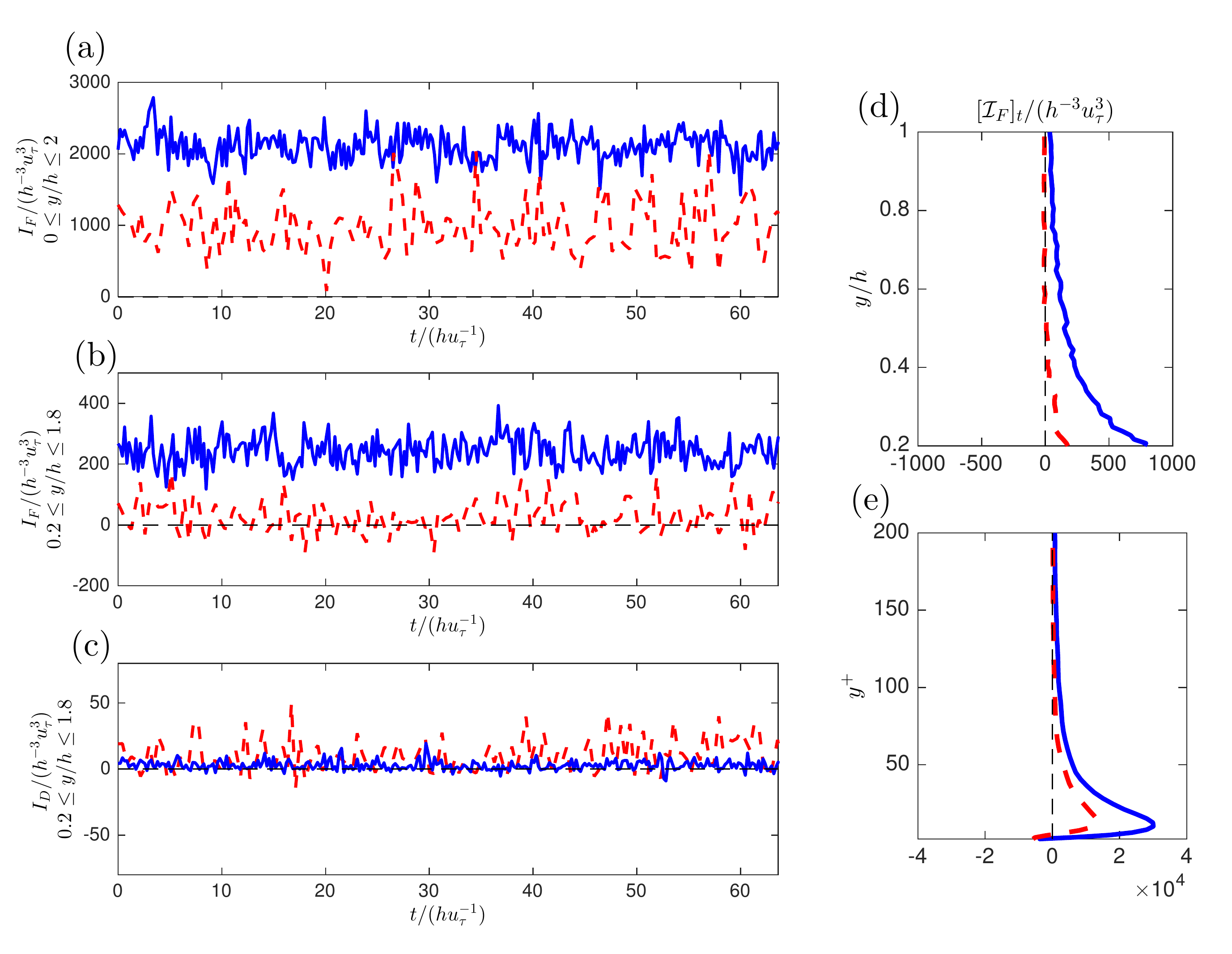}}\caption{ (a)-(c) Time series of the contribution to the time rate of change of streamwise square vorticity $\int\df y\[\Omega_x^2/2\]_z$ by perturbation torques, $I_F$, and by advection of streamwise mean vorticity by the mean flow, $I_D$, for NS940 (solid) and RNL940 (dashed). (a)~$I_F$ 
over the whole channel, $0\le y/h\le 2$ ($I_D=0$ in this case). The time mean  $I_F$  is $2103.6\,h^{-3}u_\tau^3$ for NS940 and $982.8\,u_\tau^3$ for RNL940. (b)~$I_F$ 
over the outer layer, $0.2\le y/h\le 1.8$. The  time mean $I_F$ for this region is $242.5\, h^{-3}u_\tau^3$ for NS940 and only $28.7\,h^{-3}u_\tau^3$ for RNL940. (c)~$I_D$ 
for the outer layer $0.2\le y/h\le 1.8$. The time mean $I_D$ is $2.9\, h^{-3}u_\tau^3$ for NS940 and $11.2\, h^{-3}u_\tau^3$ for RNL940.
These figures show that in NS940 and RNL940 the roll is maintained locally by the perturbation Reynolds stresses.
(d,e)~Cross-stream structure of the time averaged contribution to
the  streamwise mean vorticity generation from perturbation Reynolds stress induced torques  $[\mathcal{I}_F]_t(y)$.
\label{fig:OMx^2}}
\end{figure}

From~\eqref{eq:NSm} we have that $\Omega_x$ satisfies the equation:
\begin{align}
	\partial_t \Omega_x &=\underbrace{ -\(V\,\partial_y +W\,\partial_z \)\Omega_x  \vphantom{\(\[v'^2\]_{x}\)} }_{\textrm{D}} +   \underbrace{\(\vphantom{\[v'^2\]_{x}}\partial_{zz}-\partial_{yy}\)  \[\vphantom{v'^2}v'w'\]_{x} -\partial_{yz} \(\[w'^2\]_{x}-\[v'^2\]_{x}\)}_{\textrm{F}}+ \,\nu\,\Del \Omega_x\ .\label{eq:OMx}
\end{align}
Term~D expresses the streamwise vorticity tendency due to advection of $\Omega_x$ by  the streamwise mean flow $(V,W)$. Because there is no vortex stretching contribution to $\Omega_x$ from the  $(V,W)$ velocity field, this term only advects the $\Omega_x$ field and cannot sustain it against dissipation.
However, this term may be responsible for systematic advection of $\Omega_x$ from the inner to the outer layer. Term~F is the torque induced by the perturbation field. This is the only term that can maintain  $\Omega_x$.
The overall budget for square streamwise vorticity in the region $R$, $y_1 \le y \le y_2$, $0\le z \le L_z$, is given by:
\begin{align}
	\partial_t \, \int\limits_{y_1}^{y_2} \df y \frac1{2}\[ \vphantom{\bit\Omega_x\times\textrm{Term D}}\Omega_x^2  \]_z
	&=  \underbrace{ \vphantom{\int\limits_{y_1}^{y_2} \df y}- \frac1{2}\[ \vphantom{\bit\Omega_x\times\textrm{Term D}}\Omega_x^2\,V\]_z\left.\vphantom{\frac1{2}}\right|^{y_2}_{y=y_1} }_{=h\,I_{D}} + \underbrace{\int\limits_{y_1}^{y_2} \df y\,\[\bit\Omega_x\times\textrm{Term F}\]_z}_{=h\,I_{F}}+  { \,\nu\int\limits_{y_1}^{y_2} \df y\,\[\vphantom{\bit\Omega_x\times\textrm{Term D}}\bit\Omega_x\,\Del \Omega_x\]_z}\ ,\label{eq:OMx^2}
\end{align}
where:
\begin{equation}
	I_D (t) = h^{-1}\int_R\df y\;\mathcal{I}_{D}(y,t)  \ ,{\rm with}\ \ \mathcal{I}_{D}(y,t) =\[ \Omega_x \times \textrm{(Term~D)}\]_z\ ,
\end{equation}
is the advection into cross-stream  region, $R$, and
\begin{equation}
	I_F(t) = h^{-1}\int_R\df y\;\mathcal{I}_{F}(y,t)  \ ,{\rm with }\ \ \mathcal{I}_{F}(y,t) =\[ \Omega_x \times \textrm{(Term~F)}\]_z\ ,
\end{equation}
is the Reynolds stress torque production in region $R$.

Time series of the contributions from $I_D(t)$ and $I_F(t)$ to the $\Omega_x$ production for NS940 and RNL940, shown in figure~\ref{fig:OMx^2}a,b,c, demonstrate that $\Omega_x$ is primarily generated in situ by Reynolds stress torques. The corresponding wall-normal structure of the time mean of $\mathcal{I}_F$, representing the local contribution to streamwise mean vorticity generation from perturbation Reynolds-stress-induced 
torques,  is shown in figure~\ref{fig:OMx^2}d,e. Note that for NS940 in the outer layer the streamwise mean vorticity generation by the Reynolds stress is strongly positive at each instant.
This implies a systematic positive correlation between the roll circulation and the torque from Reynolds stress with the torque configured so as to maintain the roll.
S3T theory explains this systematic correlation between the roll/streak structure and the perturbation torques maintaining it as  a direct consequence of  the straining of the perturbation field by the streak~\citep{Farrell-Ioannou-2012}.

Having established that  the streamwise vorticity in the outer layer is maintained in situ by systematic correlation of Reynolds stress torque with the roll circulation we conclude that the SSP cycle in both NS and  RNL operates in the outer layer in a manner essentially similar to that in the inner layer.

\section{Discussion and Conclusions}

We have established that both NS and RNL produce a roll/streak structure in the outer layer and that an SSP is operating there despite the low shear in this region. It has been already shown that turbulence self-sustains in the log-layer in the absence of boundaries~\citep{Mizuno-Jimenez-2013}
 and that an SSP operates independently in the outer layer \citep{Rawat-etal-2015}. These results are consistent with our finding that an SSP
 cycle exists in both the inner-layer and outer-layer. RNL self-sustains turbulence at  moderate 
 Reynolds numbers in pressure-driven
 channel flow despite its greatly simplified dynamics when compared to NS. Remarkably, and consistent with the prediction of
 S3T that RNL turbulence is maintained by a small set of Lyapunov structures associated with the Lyapunov spectrum of the time-dependent streak, in the RNL system the turbulent state is maintained by a small set of structures with low streamwise wavenumber Fourier components (at $\Ret\approx940$ with the 
 chosen channel
 the SSP involves only the $k_x=0$ streamwise mean and  the next six
 streamwise Fourier components).  In this way RNL  produces a turbulent 
 state of reduced complexity. RNL identifies an exquisitely contrived SSP cycle which  has been previously identified to comprise the generation of the streak  through lift-up by the rolls, the maintenance of the rolls by torques induced by the perturbations which themselves are maintained by an essentially time-dependent parametric non-normal interaction with  the streak (rather than e.g.~inflectional instability of the streak structure)~\citep{Farrell-Ioannou-2012}. The vanishing of the Lyapunov exponent associated with the SSP is indicative of feedback regulation acting between the streaks and the perturbations by which the parametric instability that sustains the perturbations on the time dependent streak is reduced to zero Lyapunov exponent, so that the turbulence neither diverges nor decays.

A remarkable feature of RNL turbulence is that it is supported by a streamwise constant  SSP so that
RNL turbulence does not imply a fundamental limitation to the streamwise
extent of the streak. In a natural turbulent flow,  fluctuations may 
be expected to produce deviations from this ideal streamwise
constancy. Observations based on cross-spectral analysis determine the streamwise length of the VLSMs to be of the order of $30h$ in pipe flows and of the order of $10-15  \delta$ in boundary layer flows
\citep{Jimenez-Hoyas-2008,Hellstrom-etal-2011,Lozano-Duran-Jimenez-2014-pof}. However, \cite{Hutchins-Marusic-2007} argue that
these are underestimates of their actual length, which can in fact be arbitrarily large.
Moreover, when $k_x \ne 0$
perturbations are of large amplitude,  observation  may
suggest that a $k_x=0$ structure  has  nonzero wavenumber (cf.~\citet{Hutchins-Marusic-2007}). Consider  for example the apparent lack of $k_x=0$  structure
in figure~\ref{fig:QL_perspective}, despite
the $k_x=0$ structure of the underlying  SSP.


The centrality of streamwise constant structure to the fundamental dynamics of wall-turbulence is
consistent with predictions  of generalized stability theory \citep{Farrell-Ioannou-1996a,Farrell-Ioannou-1996b,Schmid-Henningson-2001} that both the optimal structure for growth of an initial perturbation  \citep{Butler-Farrell-1992,Farrell-Ioannou-1993a,Reddy-Henningson-1993} as well as the optimal structure for producing a response by continuous forcing \citep{Farrell-Ioannou-1996a,Bamieh-Dahleh-2001,Jovanovic-Bamieh-2005,Cossu-etal-2009,Sharma-McKeon-2013} are streamwise constant.


In this work, formation of roll/streak structures in the log-layer is attributed to the universal mechanism by which turbulence is modified by the presence of a streak in such way as to induce growth of a roll structure configured to lead to continued growth of the original streak. This growth process underlies the non-normal parametric mechanism of the SSP that maintains turbulence~\citep{Farrell-Ioannou-2012}.
This universal mechanism  neither predicts nor requires that the roll/streak
structures be of finite  streamwise extent, and in its simplest form it has been demonstrated that it supports roll/streak structures with zero streamwise wavenumber. From this point of view, the observed length of roll/streak structures is neither a primary nor necessary consequence of the SSP supporting them, but rather a secondary effect of  disruption by the  turbulence.

A distinction should be noted between mechanisms fundamentally related to streamwise constant processes and those which require that the streamwise constant
structure also be time-independent. The mechanism we have
advanced intrinsically requires that the streak be time-dependent for the parametric growth of perturbations to be supported. Because this fundamental mechanism of turbulence requires time-dependence 
of the streamwise constant streak, it predicts that turbulence must be time-dependent 
and not exclusively spatially chaotic. It further implies that mechanisms based on critical layers and modal growth processes \citep{Waleffe-1997,Hall-Sherwin-2010}
cannot support turbulence  by this mechanism
because the temporal independence of the flow required for existence of critical layers and modal instability does not obtain in these turbulent flows.

Turbulence maintained in RNL exhibits a log-layer, although with different von K\'arm\'an constants
depending on the truncation in streamwise wavenumber imposed on the RNL. {However, it should be noted that judicious choice of
the streamwise-varying components can produce von K\'arm\'an constants that are very close to those obtained in DNS (cf. \cite{Bretheim-etal-2015})}. Existence of a log-layer is a fundamental requirement of asymptotic matching between regions with different spatial scaling, as was noted by \citet{Millikan-1938}.
However, the exact value of the von K\'arm\'an constant does not have a similar fundamental basis in analysis.
RNL turbulence, which is closely related to NS turbulence but more efficient in producing Reynolds stresses, maintains  as a consequence a smaller shear and therefore greater von K\'arm\'an constant.

In this work we have provided evidence  that NS turbulence is closely related in its dynamics  to RNL turbulence from the wall through the log-layer. Moreover, given that the dynamics of RNL turbulence can be understood fundamentally from its direct relation with S3T turbulence, we conclude  that the mechanism of turbulence in wall bounded shear flow can be insightfully related to the analytically tractable roll/streak/perturbation SSP that was previously identified to maintain S3T turbulence. We conclude that the severe restriction of the 
dynamics, coupled with the restricted support of the dynamics in streamwise wavenumber that are inherent in the RNL system, result in the establishment of a statistically steady turbulent state in which, while the maintained statistics differ in particulars from those of a DNS at the same $\Re$, these systems share 
fundamental aspects of both structure and dynamics,  and that this relation provides
an attractive pathway to further understanding of wall-turbulence.

\vspace{.5em}

This work was funded in part by the Multiflow program of the European Research Council. N.C.C. was partially supported by the NOAA Climate and Global Change Postdoctoral Fellowship Program, administered by UCAR's Visiting Scientist Programs. B.F.F. was supported by  NSF \mbox{AGS-1246929}. We thank Dennice Gayme  and Vaughan Thomas
for helpful  comments.

\end{document}